\documentclass[12pt,letter,useAMS,natbib]{mn2e}
\usepackage{graphicx,amssymb}
\usepackage{txfonts}
\usepackage{natbib_jrm}

\bibpunct[, ]{(}{)}{;}{a}{}{,}
\def \aj {AJ}
\def \mnras {MNRAS}
\def \pasp {PASP}
\def \apj {ApJ}
\def \apjs {ApJS}
\def \apjl {ApJL}
\def \aap {A\&A}
\def \aaps {A\&AS}

\def \araa {ARAA}
\def \iaucirc {IAUC}

\def\lesssim{\mathrel{\hbox{\rlap{\hbox{\lower4pt\hbox{$\sim$}}}\hbox{$<$}}}}
\def\gtrsim{\mathrel{\hbox{\rlap{\hbox{\lower4pt\hbox{$\sim$}}}\hbox{$>$}}}}

\long\def\symbolfootnote[#1]#2{\begingroup%
\def\thefootnote{\fnsymbol{footnote}}\footnote[#1]{#2}\endgroup}

\begin{document}


\title[Late-time view of progenitors of Type IIP supernovae]{A Late-Time View of the Progenitors of Five Type IIP Supernovae}
\author[Maund et al.]{Justyn R. Maund$^{1,2,3}$\thanks{Email: j.maund@qub.ac.uk}, Emma Reilly$^{1}$ and Seppo Mattila$^{4}$\\
$^{1}\,$Department of Physics and Astronomy, Queen's University, Belfast BT7 1NN, Northern Ireland, UK\\
$^{2}\,$Dark Cosmology Centre, Niels Bohr Institute, University of Copenhagen, Juliane Maries Vej 30, 2100 Copenhagen, DK.\\
$^{3}\,$Royal Society Research Fellow\\
$^{4}\,$Finnish Centre for Astronomy with ESO (FINCA), University of Turku,
V\"ais\"al\"antie 20, FI-21500 Piikki\"o, Finland
}

\maketitle
\begin{abstract}
The acquisition of late-time imaging is an important step in the analysis of pre-explosion observations of the progenitors  of supernovae.  We present late-time HST ACS WFC observations of the sites of five Type IIP SNe: 1999ev, 2003gd, 2004A, 2005cs and 2006my.  Observations were conducted using the $F435W$, $F555W$ and $F814W$ filters.  We confirm the progenitor identifications for SNe 2003gd, 2004A and 2005cs, through their disappearance.  We find that a source previously excluded as being the progenitor of SN~2006my has now disappeared.  The late-time observations of the site of SN~1999ev cast significant doubt over the nature of the source previously identified as the progenitor in pre-explosion WFPC2 images.  The use of image subtraction techniques yields improved precision over photometry conducted on just the pre-explosion images alone.  In particular, we note the increased depth of detection limits derived on pre-explosion frames in conjunction with late-time images.
We use SED fitting techniques to explore the effect of different reddening components towards the progenitors.  For SNe 2003gd and 2005cs, the pre-explosion observations are sufficiently constraining that only limited amounts of dust (either interstellar or circumstellar) are permitted.  Assuming only a Galactic reddening law, we determine the initial masses for the progenitors of SNe  2003gd, 2004A, 2005cs and 2006my of $8.4\pm2.0$, $12.0\pm2.1$, $9.5^{+3.4}_{-2.2}$ and $9.8\pm1.7M_{\odot}$, respectively.
\end{abstract}
\section{Introduction}
\label{sec:intro}
All stars with initial masses $>8M_{\odot}$ are expected to end their
lives as core-collapse supernovae (CCSNe).  In the last decade, the
direct observation of the progenitors of CCSNe in fortuitous
pre-explosion imaging, has become an integral step in the study of all
nearby events \citep[for a review see][]{2009ARA&A..47...63S}.

The majority of the success in the actual detection of progenitors has
been for the Red Supergiant (RSG) precursors to hydrogen-rich Type II
Plateau (IIP) SNe.  The ensemble of progenitor detections and
detection limits for these RSGs, however, presented a conflict with
the theoretical expectation from stellar evolution models
\citep{2008arXiv0809.0403S}.  The ``Red Supergiant Problem'' refers to
the apparent absence of progenitors of Type IIP SNe with mass
$>17M_{\odot}$, whereas the theoretical expectation is that stars with
masses up to $25-35M_{\odot}$ should end their lives as RSGs.
Recently \citet{2011MNRAS.412.1522S} and \citet{2012MNRAS.419.2054W}
considered the role of circumstellar dust, which is otherwise not probed
by the resulting SN or the surrounding stellar population, as a
possible solution to the RSG problem.  In particular,
\citeauthor{2012MNRAS.419.2054W} established that the amount of
circumstellar dust, and hence reddening, is larger for higher
luminosity RSGs, that arise at the upper end of the mass range for
stars to explode as Type IIP SNe.  In the case of SN~2012aw,
\citet{2012arXiv1204.1523F} and \citet{2012ApJ...756..131V} observed a
significant difference between the reddening determined towards the
progenitor and the reddening inferred towards the subsequent SN;
suggesting a significant amount of dust was associated with the
progenitor and that it was destroyed in the SN explosion, such that it
could not be measured post-explosion.  \citet{2012ApJ...759...20K} and
\citet{2012ApJ...756..131V} also considered the additional possibility
that the nature of the reddening, due to a circumstellar dust
component, was different to ordinary reddening laws appropriate for
the interstellar medium \citep[e.g.][]{ccm89}.

A further issue with the RSG problem is that the maximum mass is
poorly constrained, due to the paucity of progenitors with high
inferred masses.  This may reflect either a real deficit of high mass
progenitors (i.e. that the upper mass limit for stars to explode as
Type IIP SNe is low) or may be due to low numbers of high mass
progenitors according to the Initial Mass Function (IMF).  The exact
articulation and quantification of the RSG problem is compounded by
the reliance on, generally, poor fortuitous pre-explosion images,
limited detections in pre-explosion images in multiple filters, and
ambiguity as to whether the object observed at the SN position is
indeed the progenitor object.

\citet{2009Sci...324..486M} demonstrated that it is possible, using
deep late-time images that are tailored (in a way that the
pre-explosion images, by their fortuitous nature, cannot be) to
observe the site of a given SN and overcome the severe limitations of
analysis on just the pre-explosion images alone.  Using image
subtraction techniques, such as {\sc isis}
\citep{1998ApJ...503..325A,2000A&AS..144..363A}, the late-time images
can be used as templates, to both demonstrate the disappearance of the
progenitor candidate (confirm the original identification) and conduct
precise photometry (without contamination from nearby and underlying
objects).

Here we present a new analysis of five previously identified
progenitor candidates using newly acquired late-time HST Advance
Camera for Surveys (ACS) Wide Field Channel (WFC) imaging.  These
progenitors are for SNe 1999ev, 2003gd, 2004A, 2005cs and 2006my, and
details of these SNe are presented in Table \ref{tab:obs:sne}.

\section{Observations}
\label{sec:obs}
The analysis of the pre-explosion HST observations of the sites of the
five target SN progenitors has been previously presented by
\citet{2005astro.ph..1323M} (1999ev); \citet{2003PASP..115.1289V},
\citet{smartt03gd} and \citep{2009Sci...324..486M} (2003gd);
\citet{2006MNRAS.369.1303H} (2004A); \citet{2005MNRAS.364L..33M} and
\citet{2005astro.ph..7394L} (2005cs); and \citet{2007ApJ...661.1013L},
\citet{2008PASP..120.1259L} and \citet{2011MNRAS.410.2767C} (2006my).
The pre- and post-explosion and late-time observations of the sites of
the five target Type IIP SN progenitors are presented in Table \ref{tab:obs}.
\begin{table*}
\caption{\label{tab:obs:sne} Details of the five target Type II SNe, with progenitor candidates identified in pre-explosion HST images.}
\begin{tabular}{lclccc}
\hline
Supernova & Type   & Host Galaxy   & Distance$^{1}$       & $E(B-V)_{gal}^{2}$  & $[O/H]^{3}$\\
          &        &               & (Mpc)          &           & (dex)    \\
\hline
1999ev    & II(P)  & NGC 4274      & $15.1\pm2.6$   & 0.020       &  8.5\\
2003gd    & IIP    & NGC 628 (M74) & $9.3\pm1.8$  & 0.061    &   8.4  \\
2004A     & IIP    & NGC 6207      & $20.3\pm 3.4$ & 0.014     &   8.3 \\
2005cs    & IIP    & NGC 5194 (M51 & $8.4\pm1.0$ & 0.031 &    8.7    \\
2006my    & IIP    & NGC 4651      & $22.3\pm2.6$     & 0.024     &    8.7\\
\hline\hline
\multicolumn{6}{l}{$^{1}\,$ After \citet{2008arXiv0809.0403S}.}\\
\multicolumn{6}{l}{$^{2}\,$ After \citet{2011ApJ...737..103S}, as quoted by NED.}\\
\multicolumn{6}{l}{$^{3}$\citet{2008arXiv0809.0403S}.}\\
\end{tabular}
\end{table*}
\subsection{Pre-explosion observations}
\label{sec:obs:pre}
Due to the availability of more recent and appropriate calibrations,
new versions of the pre-explosion WFPC2 and ACS data were retrieved
from the HST archive\footnote{http://archive.stsci.edu}.  The
pre-explosion observations acquired with WFPC2 were drizzled together
following the standard
procedure\footnote{http://www.stsci.edu/hst/wfpc2/analysis/WFPC2\_drizzle.html}.
These images were produced for the purposes of image subtraction (see
Section \ref{sec:obs:sub}).  In parallel, a separate ``reduction'' and
photometric analysis was conducted on the same data using HSTphot \citep{dolphhstphot}.  For SN~2005cs, pre-explosion images
were acquired with the Advanced Camera for Surveys (ACS) Wide-Field
Channel (WFC).  These images were acquired using a four-point box
dither pattern.  Alignment between the images was checked using the
{\sc PyRAF} task {\it tweakshifts}, and the images were combined using
{\it multidrizzle}.  Due to the half-integer pixel shifts of the
dither pattern it was possible to enhance the spatial sampling of the
final image, providing a final pixel scale of $0.035\arcsec$.  This is
larger than exact half-sampling ($0.025\arcsec \, \mathrm{px^{-1}}$),
but is used to match the sampling achieved for the late-time ACS WFC
images of SN2005cs (see Section \ref{sec:obs:late}).
\subsection{Post-explosion observations}
\label{sec:obs:post}
For the purposes of this study, the principal interest in the
immediate post-explosion images (acquired up to 3 years
post-explosion) was to provide a position for the SN relative to the
surrounding stars, such that the SN position could be identified on
the pre-explosion and late-time frames through differential
astrometry.  The ``reduction'' procedure for these images was the same
as outlined for the pre-explosion observations (see Section
\ref{sec:obs:pre}).
\subsection{Late-time ACS observations}
\label{sec:obs:late}
Late-time observations of the sites of four of the target Type IIP SNe
were acquired using the HST ACS WFC 1 for the program GO-11675 (PI
Maund).  The WFC chip was windowed to an array of $\mathrm{1k \times
  1k}$ pixels to reduce the readout time and mitigate the role of
Charge Transfer Inefficiency (CTI).  Observations were conducted in
three filters $F435W$, $F555W$ and $F814W$.  Importantly, for two SNe
(2003gd and 2004A) the use of the $F555W$ filter in these late-time
observations {\it does not} match the pre-explosion images acquired
with the wider $F606W$ filter.  In the pre-explosion frames, the
majority of the flux from the progenitor at these wavelengths is
representative of the continuum (requiring a small $F555W-F606W$
colour correction).  At late-times, however, the $F606W$ filter
encompasses the wavelength of $H\alpha$, which is a characteristic
emission feature of late-time Type IIP SN spectra.  A more appropriate
comparison between before and after continuum fluxes is, therefore,
achieved with the $F555W$ filter although, as discussed below,
corrections for the slightly different filter transmission functions
also need to be considered.  Each of these late-time observations is
composed of four separate sub-exposures acquired in a 4-point box
dither pattern.  This arrangement was used to permit the acquisition
of better spatial sampling of the point-spread function (PSF) and for
removal of fixed hot-pixel features.  The {\sc PyRAF} task {\it
  multidrizzle} was used to drizzle each of the sub-exposures (for a
given filter) together.  The task {\it tweakshifts} was used to
fine-tune the alignment between each of the sub-exposures prior to
drizzling.  Although the 4-point box dither pattern employs
half-integer pixel shifts, potentially giving an improvement of
spatial sampling by a factor of 2, the presence of aliasing
(alternating bands across the frames) prohibited reaching this final
pixel-scale.  This phenomenon was relatively insensitive to the choice
of the drizzling kernel.  Instead the final pixel scale is greater
than one half of the original $0.05\arcsec$ pixel scales of ACS/WFC
($0.035\arcsec\,\mathrm{px^{-1}}$).

For SN 2006my, observed for program GO-12282 (P.I. D. Leonard), the
late-time images were only acquired in two bands ($F555W$ and
$F814W$).  For each filter, two exposures were acquired, for the
rejection of cosmic rays, but at the same pointing such that no
spatial resampling was possible.  These images for each filter were
combined using {\it multidrizzle}, but with the output images having
the original ACS/WFC pixel scale ($0.05\arcsec\,\mathrm{px^{-1}}$).
\subsection{Geometric Transformations}
\label{sec:obs:geo}
A series of transformations were calculated for the complete datasets
to determine the positions for objects in a common reference frame.
Geometric transformations were calculated between the pre-,
post-explosion and late-time $F555W$ images (or, if unavailable,
$F814W$ images) using the {\sc iraf} task {\it geomap}, assuming only
simple offsets, rotations and scalings.  For the data at a given
epoch, shifts between images with other filters and the corresponding
reference $F555W$ image were calculated by cross-correlating these
images with the $F555W$ image as the reference frame. The
cross-correlation was facilitated using the {\sc PyRAF} task {\it
  crosscor}, with the corresponding shifts calculated using {\it
  shiftfind}.
\subsection{Photometry}

For WFPC2 pre-explosion images the HSTphot package
\citep{dolphhstphot} was utilised to conduct PSF-fitting photometry of
the input images. HSTphot provides the appropriate corrections
for aperture size and charge transfer inefficiency, as well as tools
for conducting artificial star tests.  We note that HSTphot
only provides aperture corrections to a final aperture size of
$0.5\arcsec$.  Using the corrections of \citet{holsper95}, we apply a
term to correct to the photometry to an infinite aperture.

For data acquired for program GO-11675 (PI Maund), principal
photometry was conducted using {\sc iraf} DAOphot on the final
output drizzled images with the enhanced spatial sampling.  We used
the latest zeropoints appropriate for ACS
WFC\footnote{http://www.stsci.edu/hst/acs/analysis/zeropoints/\#tablestart}.
Aperture corrections were calculated for each image to an aperture of
$0.5\arcsec$, with a further correction to infinity adopted from
\citet{acscoltran}.

A key concern for the fidelity of the derived photometry is the
inefficiency of charge transfer (CTI) for charged coupled detectors on
HST.  As photometry was conducted on the drizzled, subsampled images,
evaluating the magnitude of the CTI on the final photometry is
non-trivial and is based on the position of a given star on the
original undrizzled, distorted {\it FLT} images.  Geometric
transformations were calculated between the final drizzled images and
the $FLT$ images, using 3rd order polynomials in $x$ and $y$, using
{\it geomap}.  This approach was used, over using a simple
pre-computed distortion table, as non-negligible shifts were found
between the expected pointings in the dither pattern.  The positions of stars
on the output drizzled images were transformed to the corresponding
locations on each of the four constituent {\it FLT} images.  Following
\citet{2008AJ....135.1900A}, we conducted small aperture (3px)
photometry on the individual {\it FLT} images and used the measured
flux and sky background values to calculate the magnitude loss due to
CTI following the analytic prescription of
\citet{2009acs..rept....1C}.  Prior to conducting aperture photometry,
the {\it FLT} images were scaled with the corresponding Pixel Area Map
\footnote{http://www.stsci.edu/hst/acs/analysis/PAMS}.  For a given
star it might be only possible to calculate the magnitude loss for 1
or 2 of the input {\it FLT } images, because of the relative positions
of bad pixels or cosmic rays.  An average CTI loss (as a magnitude)
was determined over the four constituent {\it FLT }images and applied
to the photometry derived from the drizzled images.

In addition, photometry of the ACS images was also conducted using the
DOLPHOT package \footnote{http://americano.dolphinsim.com/dolphot/}.  We
utilised two implementations of DOLPHOT for photometry of the
ACS data: {\sc DOLphot} with the ACS module on the distorted {\it CRJ}
and {\it FLT} frames (which we refer to as DOLPHOT/{\it ACS})
and DOLPHOT as a generic photometry package for the
distortion-corrected drizzled frames.  We find excellent agreement
between our DAOphot photometry and the photometry derived using
HSTphot and DOLPHOT, within the limits of the photometric
uncertainties.  Similarly to HSTphot, the ACS photometry was
corrected from a $0.5\arcsec$ to an infinite aperture, using the
corrections tabulated by \citet{acscoltran}

The data for SN 2006my were analysed separately using only the
  HSTphot and DOLPHOT packages for the WFPC2 and ACS data,
respectively.

\begin{table*}
\caption{\label{tab:obs}HST observations of the sites of the five
  target Type II SNe.}
\begin{tabular}{lllccccc}
\hline\hline
 & Dataset & Date & Instrument & Filter & Exposure  & Final Pixel & Program \\
  &      &      &            &        &  Time (s) &    Size ($\arcsec$)        &         \\
\hline
\multicolumn{8}{c}{{\bf SN 1999ev}}\\
Pre-explosion & U2JF0101/02T/03T & 1995 Feb 5 & WFPC2/WF2 & F555W  & 280 & 0.1  & 5741$^{1}$    \\
\\
Post-explosion&J8DT03010 & 2001 Dec 31 & ACS/WFC1 & F555W & 450 & 0.05 & 9353$^{2}$ \\
              &J8DT03020 & 2001 Dec 31 & ACS/WFC1 & F814W & 450 & 0.05 & 9353 \\
              &J8DT03030 & 2001 Dec 31 & ACS/WFC1 & F435W & 400 & 0.05 & 9353 \\
\\
Late-time     &JB4T01010 & 2010 Nov 14 & ACS/WFC1 & F555W & 1368 & 0.035 & 11675$^{3}$ \\
              &JB4T01020 & 2010 Nov 14 & ACS/WFC1 & F814W & 1408 & 0.035 & 11675 \\ 
              &JB4T01030 & 2010 Nov 14 & ACS/WFC1 & F435W & 1608 & 0.035 & 11675 \\
\hline
\multicolumn{8}{c}{{\bf SN 2003gd}}\\
Pre-explosion &U8IXCA01M/02M & 2002 Aug 25 & WFPC2/WF2 & F606W & 1000 & 0.1 & 9676$^{6}$ \\
              &U8IXCY01M/02M/03M & 2002 Aug 28 & WFPC2/WF2 & F606W & 2100 & 0.1 & 9676 \\
\\
Post-explosion&J8NV01020 & 2003 Aug 1 & ACS/HRC & F435W & 2200 & 0.025 & 9733$^{5}$ \\
              &J8NV01040 & 2003 Aug 1 & ACS/HRC & F555W & 1000 & 0.025 & 9733 \\
              &J8NV01050 & 2003 Aug 1 & ACS/HRC & F814W & 1350 & 0.025 & 9733 \\
\\
Late-time     &JB4T02010 & 2010 Nov 14 & ACS/WFC    & F555W  & 1364      & 0.035     & 11675$^{3}$   \\
              &JB4T02020 & 2010 Nov 14 & ACS/WFC    & F814W  & 1398      & 0.035     & 11675   \\
              &JB4T02030 & 2010 Nov 14 & ACS/WFC    & F435W  & 1600      & 0.035     & 11675   \\
\hline
\multicolumn{8}{c}{{\bf SN 2004A}}\\
Pre-explosion &U6EAD001R/02R & 2001 Jul 2  & WFPC2/WF3  & F814W & 460  & 0.1 & 9042$^{4}$\\
              &U6EAD003R/04R & 2001 Jul 2  & WFPC2/WF3  & F606W & 460 & 0.1& 9042\\
              \\
Post-explosion&J8NV03010 & 2004 Sep 23 & ACS/WFC1 & F435W & 1400 & 0.05 & 9733$^{5}$ \\
              &J8NV03020 & 2004 Sep 23 & ACS/WFC1 & F555W & 1509 & 0.05 & 9733 \\
              &J8NV03030 & 2004 Sep 23 & ACS/WFC1 & F814W & 1360 & 0.05 & 9733 \\
\\
Late-time     &JB4T03010 & 2010 Sep 09 & ACS/WFC & F555W & 1400 & 0.035 & 11675$^{3}$ \\
              &JB4T03020 & 2010 Sep 09 & ACS/WFC & F814W & 1434 & 0.035 & 11675 \\
              &JB4T03030 & 2010 Sep 09 & ACS/WFC & F435W & 1636 & 0.035 & 11675 \\

\hline
\multicolumn{8}{c}{{\bf SN 2005cs}}\\
Pre-explosion     &J97C5     & 2005 Jan 20-21 & ACS/WFC  & F435W & 2720 & 0.035 & 10452$^{7}$ \\
              &J97C5     & 2005 Jan 20-21 & ACS/WFC  & F555W & 1360 & 0.035 & 10452 \\
              &J97C5     & 2005 Jan 20-21 & ACS/WFC  & F658N & 2720 & 0.035 & 10452 \\
              &J97C5     & 2005 Jan 20-21 & ACS/WFC  & F814W & 1360 & 0.035 & 10452 \\

\\
Post-explosion&J9AR01011-31&2005 Jul 24 & ACS/HRC   & F555W & 1944 & 0.025 & 11675 \\
\\
Late-time     &JB4T04010 & 2010 Jul 30 & ACS/WFC1    & F555W & 1460 & 0.035 & 11675 \\
              &JB4T04020 & 2010 Jul 30 & ACS/WFC1    & F814W & 1494 & 0.035 & 11675 \\
              &JB4T04030 & 2010 Jul 30 & ACS/WFC1    & F435W & 1696 & 0.035 & 11675 \\
              \hline
\multicolumn{8}{c}{{\bf SN 2006my}}\\
Pre-explosion & U2DT0901T/02T/03T & 1994 May 20 & WFPC2/WF2 & F555W & 660 & 0.1 & 5375$^{10}$ \\
 & U2DT0904T/05T/06T & 1994 May 20 & WFPC2/WF2 & F814W & 660 & 0.1 & 5375 \\
\\
Post-explosion &U9OX0301M/02M/03M/04M & 26 Apr 2007 & WFPC2/PC & F555W & 1200 & 0.05 & 10803$^{11}$\\
 &U9OX0305M/06M & 26 Apr 2007 & WFPC2/PC & F814W & 1200 & 0.05 & 10803\\
  &U9OX0307M/08M & 26 Apr 2007 & WFPC2/PC & F450W & 1400 & 0.05 & 10803\\
\\
Late-time&JBKS01010& 21 Nov 2010 & ACS/WFC & F555W & 1090 & 0.05 & 12282$^{12}$\\
&JBKS01020& 21 Nov 2010 & ACS/WFC & F814W & 1090 & 0.05 & 12282\\
\hline\hline
\end{tabular}
\\
\begin{tabular}{llll}
$^{1}$ P.I. J. Westphal & $^{2}$ P.I. S.J. Smartt & $^{3}$ P.I. J.R. Maund & $^{4}$ P.I. S.J. Smartt\\
$^{5}$ P.I. S.J. Smartt & $^{6}$ P.I. J. Rhoads & $^{7}$ P.I. S. Beckwith & $^{8}$ P.I. Filippenko\\
$^{9}$ P.I. M. Meixner & $^{10}$ P.I. V. Rubin & $^{11}$ P.I. S.J. Smartt & $^{12}$ P.I. D. Leonard\\
\hline
\end{tabular}
\end{table*}

\subsection{Image Subtraction}
\label{sec:obs:sub}
Image subtraction techniques were used to conduct template subtraction
of the late-time images from the pre-explosion images to: 1) confirm
the identities of the progenitors through disappearance; and 2)
conduct optimal differential photometry, independently of the
background, of the now absent progenitors.  We utilised the {\sc isis
  v2.2} image subtraction package
\citep{1998ApJ...503..325A,2000A&AS..144..363A}, which matches the
PSFs of the input and template/reference images (as well as refining
the alignment between the two images and scaling the flux levels).  In
addition, {\sc isis} also provides automatic object detection and
photometry on the difference images.  We also tested our image
subtractions using the HOTPANTS image subtraction
package\footnote{http://www.astro.washington.edu/users/becker/hotpants.html},
and examined the resulting difference images using DAOphot.  We
found no systematic difference between the photometry of difference
images calculated using the two packages, and for this study use
photometry derived from difference images constructed using {\sc isis}

For each SN the late-time images were, generally, used as the
reference images.  The late-time images were transformed to match the
pre-explosion images using {\sc iraf} task {\it geotran}.  This
ensured that we avoided resampling the lower quality pre-explosion
images to match the superior late-time images.

The evaluation of the systematic uncertainties was conducted by
varying the key parameters in {\sc ISIS} that principally affected the
output photometry: the number and size of the stamps used for
calculating the kernel, the degree of the kernel variation across the
field and the degree of the background fitting function.  A major
concern for conducting image subtraction analysis using HST images is
the effect of the degree by which the PSF is subsampled.  Subsampling
of the PSF means that the majority of the flux in {\it WFPC2} images
will fall in a single pixel, which may induce systematic errors in the
construction of the convolution kernel.  In order to assess the
systematic uncertainty associated with the degree of subsampling,
iterations of the image subtraction routine were conducted with
different degrees of Gaussian smoothing applied to the input and/or
reference images.  We conservatively estimate that the total
systematic uncertainties associated with the image subtraction process
are $\sim 5-10\%$ of the flux observed in the difference image (as
discussed in Section \ref{sec:res}; with the larger systematic
uncertainty associated with the fainter progenitors, in addition to
the commensurate increase in the relative Poisson noise).  The
systematic uncertainties dominate over the Poisson noise, and we
assume dominate over other noise sources, such as read noise, that are
propagated to the difference images.

As we are concerned with pre-explosion and late-time images acquired
with HST, the image subtraction process will involve CTI in both the
input and reference images.  There are two keys issues with the
evaluating the effect of CTI on photometry that has been derived using
difference imaging: 1) the CTI that affects the progenitor flux is in
the flux system of the pre-explosion image, whereas the photometry of
the progenitor from the difference image is found in the flux system
of the late-time reference image; and 2) the evaluation of CTI for a
source using its photometry from the pre-explosion image explicitly
undermines any increase in the precision of photometry that might be
derived using image subtraction techniques.

For both WFPC2 and ACS/WFC the CTI is principally dependent on the
flux of the object, the level of the nearby background, the position
of the object on the chip (the number of charge transfers to be made
to readout the electrons) and the date at which the observations were
made.  Rather than attempting to explicitly calculate the counts
associated with the progenitor on the pre-explosion image, we instead
use photometry of nearby or artificial stars, in the vicinity of the
progenitor, to serve as proxies for the calculation of the CTI for the
progenitor.  As these stars are at approximately the same position on
the chip, on similar backgrounds as the progenitor and observed at the
same epoch, we can reduce the problem of determining the CTI to just
the dependence on the brightness (or magnitude) of the progenitor.
The nearby stars (real or fake) sample a range of brightnesses and
simple expressions can be derived relating the magnitude of an object
in the difference image to the CTI (in magnitudes) directly.

We can consider the flux measured on the images, uncorrected for CTI,
as $f^\prime$.  The flux corrected for CTI is then simply:
\begin{equation}
f=f^\prime10^{-0.4\,CTI}
\end{equation}
Under the assumption that the CTI is a relatively small effect, we use the
initial approximation that $m \approx m^{\prime}$ to derive the CTI in the
pre-explosion frame using photometry, derived from difference images,
in the photometric system of the pre-explosion images.

On the ACS images, we model the CTI as being effectively dependent on
only two parameters: the brightness of the object and its $y$-position
on the {\it FLT} images.  We established, for the images considered
here, that the effect of nearby background inhomogeneities, around the
progenitor positions, are well within the stated uncertainties of the
CTI expressions.  We consider the CTI for a given range of pixels to be
approximately given by a power law dependent only on the magnitude of
the object:
\begin{equation}
CTI(m)=\beta m ^{\alpha}	
\label{eqn:obs:ctiacs}
\end{equation}
We evaluated the coefficients of Equation \ref{eqn:obs:ctiacs} for
each pre-explosion ACS/WFC image (as the coefficients are dependent on
the date on which the observations were made and the specific
background at the progenitor position), using real and artificial
stars for which the CTI had been evaluated using the equations of
\citet{2009acs..rept....1C}.  Due to significant uncertainty in the
expression used to determine the CTI, and its slowing varying nature
with pixel position, it is possible to consider the CTI to be
approximately fixed over $\sim 50$ pixel ranges in $y$. 

We make a similar approximation for the WFPC2 observations, using the
CTI formulation presented by \citet{dolp00cte}\footnote{with updates
  from $\mathrm{http://purcell.as.arizona.edu/wfpc2\_calib/}$},
derived using artificial stars generated using HSTphot.  We find
the average dependence of the CTI (over a 50 pixel range in
$y$-position) to correspond to a second order polynomial:

\begin{equation}
CTI(m)=\alpha+\beta m +\gamma m^{2} 
\label{eqn:obs:ctiwfpc2}
\end{equation}

For both $WFPC2$ and $ACS$ observations, the dependence on $m$ is
relatively weak; the difference in its evaluation using $m$ or
$m^{\prime}$ is negligible.  The importance of this approach is that
it avoids specifically determining fluxes and sky background values from
the pre-explosion images (which defeats both the purpose and precision
afforded by using image subtraction techniques to derive the
photometry of the pre-explosion source).  The zeropoint in the
photometric scale of the reference image $Z_{R}$ can be derived from
photometry $m_{R}$ of reference stars, identified by {\sc isis}.
Using a package such as {\sc isis}, $f^{\prime}_{R}$ of the reference
objects in the reference image can be measured directly using aperture
photometry and can be compared directly with the photometry of the
same stars derived using DAOphot, DOLPHOT or HSTphot; such that $Z_{R}$ contains not only the absolute
zeropoint, but also all the relevant aperture corrections.  The
magnitude $m_{d}$ of the progenitor candidate can then be found
directly.  As the image subtraction procedure determines the
difference in observed fluxes $f^{\prime}$, these fluxes must be
further corrected for CTI derived on the input image (containing the
progenitor) using the scheme outlined above.  Given the difference
measured from reference and input images, the final magnitude of the
pre-explosion source is given as:
\begin{equation}
m_{d}=-2.5 \log_{10}\left( f^{\prime}_{d} \right) + Z_{R} + CTI_{i}
\end{equation}

\begin{figure}
\includegraphics[width=8cm]{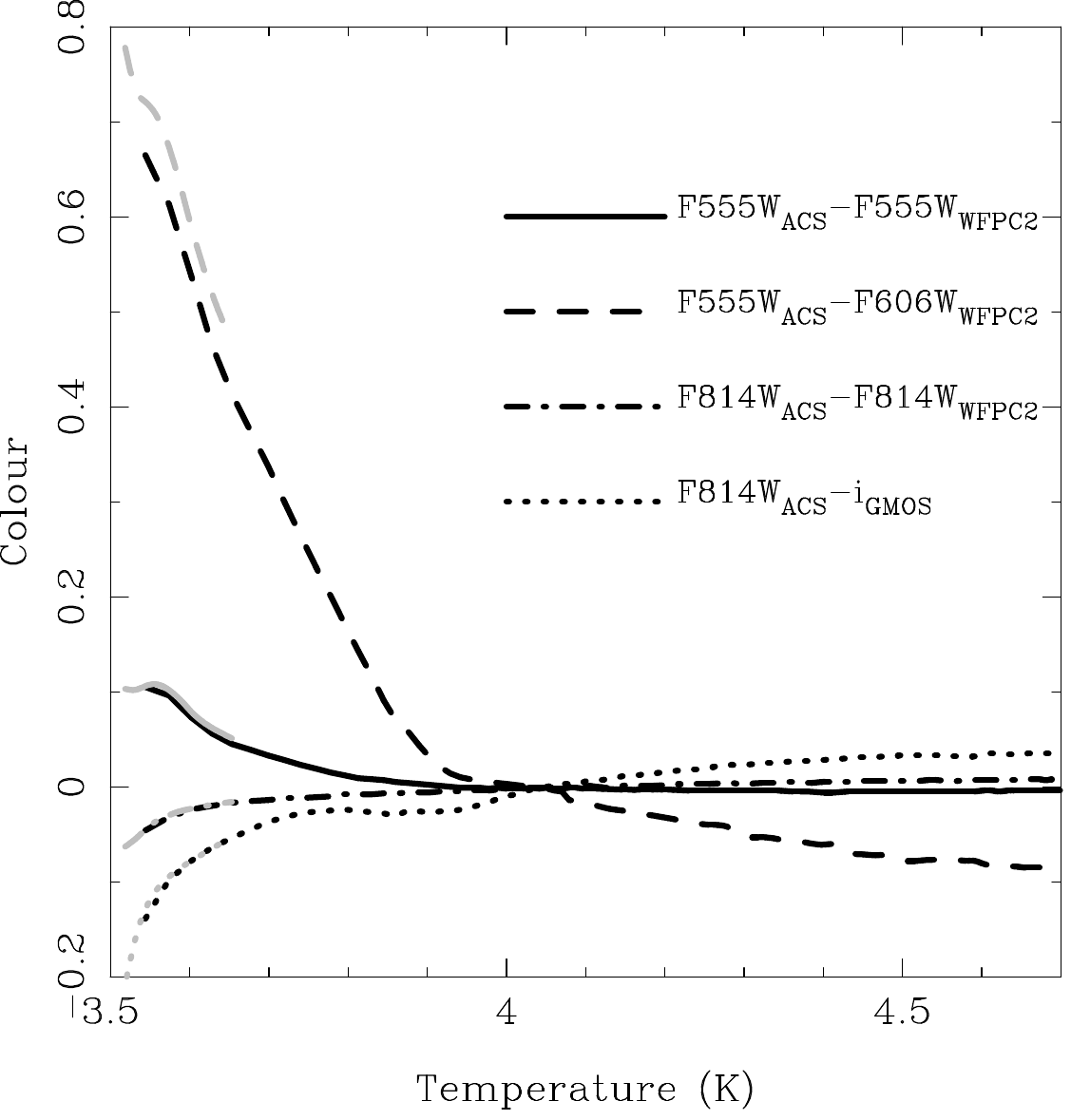}
\caption{Colour terms, between filter sets, as a function of temperature for ATLAS9 (heavy lines; \citealt{2004astro.ph..5087C}) and MARCS (grey line; \citealt{marcsref}) model SEDs (appropriate for supergiants with $E(B-V)=0$; see Section \ref{sec:ana}).}
\label{fig:obs:colcor}
\end{figure}

An additional source of systematic uncertainty is the differences in
the filter transmission functions between images used for image
subtraction analysis.  We note that, although some filters are
nominally identical, there may also be differences between the same
filters used on different instruments.  We used synthetic photometry
of ATLAS9 \citep{2004astro.ph..5087C} and MARCS \citep{marcsref} model
SEDs, using the total transmission (filter and instrument) functions,
to determine the relative colour differences between the filter sets
used here.  For most combinations of filters, in particular between
nominally identical HST filters, the colour difference as a function
of temperature is $<0.1$ mags for low reddening (see
Fig. \ref{fig:obs:colcor}).

\subsection{Non-detections and detection limits}
\label{sec:obs:lim}
In previous studies
\citep[e.g.][]{2005astro.ph..1323M,2011MNRAS.410.2767C}, the
derivation of the detection thresholds has been conducted using
analytical expressions for the background and source noise for an
``ideal'' observation.  This approach, however, does not accurately
reflect the way in which stars are actually detected in the photometry
process, using {\sc iraf} tasks such as {\sc DAOfind}, and the effect
of crowding.  We consider the insertion and attempted recovery of
artificial stars to derive the detection threshold.

Artificial stars were generated using the PSFs derived from the data
themselves, with a randomly selected magnitude from a uniform
distribution and with a position uniformly distributed within $\pm
0.5\,\mathrm{px}$ in $x$ and $y$ of the SN location.  The first
approach involved repeating the original detection and aperture and
PSF photometry routine on the pre-explosion images, with artificial
stars inserted, and considering a detection to be any recovery of a
star within 1 pixel and 0.5 magnitudes of the input star's parameters.
The second approach utilised {\sc isis} to conduct image subtraction between
the late-time images and pre-explosion images, in which artificial
stars had been inserted in the latter.  For this latter approach, we
set the coordinates at which {\sc isis} was to conduct aperture photometry
and classified a detection to be any instance in which the recovered
flux was 3 times that of the corresponding noise (including the
systematic uncertainty; see section \ref{sec:obs:sub}).

We consider the detection threshold to be the magnitude at which we
recover $50\%$ of input artificial stars, using a $3\sigma$ detection
threshold with DAOphot.  As noted by Maund (2013, in prep.), the
completeness function can be considered in terms of the complementary
cumulative Gaussian distribution.  We therefore quote the
corresponding width of the completeness function as an effective
uncertainty on the derived detection threshold.

\section{Observational Results}
\label{sec:res}
\subsection{SN 1999ev}
\label{sec:res:99ev}
SN~1999ev was discovered by T. Boles \citep{99eviauc1} on 1999 Nov
7.225 in the galaxy NGC~4274.  \citet{99eviauc1a} subsequently
classified the SN as being of Type II, although no further
sub-classification of the SN has been reported.  \citet{vandykprog}
attempted to identify the progenitor object in pre-explosion WFPC2
$F555W$ images from 1995 Feb 1, although were not able to conclusively
identify a single object as the progenitor.  In an independent
analysis, using a differential astrometric solution derived using
post-explosion ACS WFC images containing the SN,
\citet{2005astro.ph..1323M} were able to identify a star in the
pre-explosion images coincident with the SN position (with an
uncertainty of $0.02\arcsec$).

\begin{figure}
\includegraphics[width=8cm]{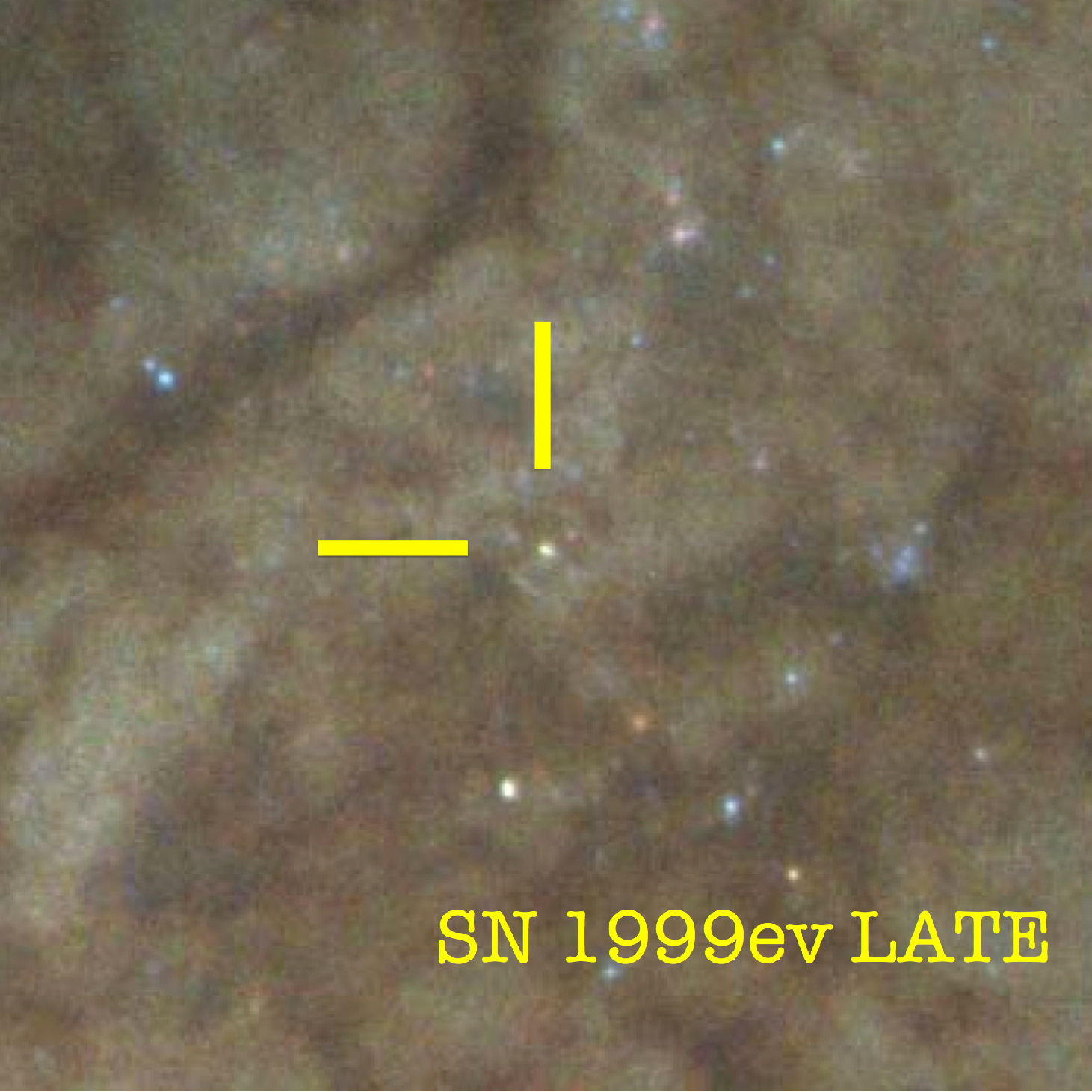}
\caption{Late-time colour image (composed of $F435W$, $F555W$ and
  $F814W$ images) of the area of NGC~4274 containing
  SN~1999ev.  The image has dimensions $15\arcsec \times 15\arcsec$,
  and is oriented such that North is up, and East is left. The
  position of SN~1999ev is located at the centre of the image.}
\label{fig:obs:99evcol}
\end{figure}

\begin{figure*}
\includegraphics{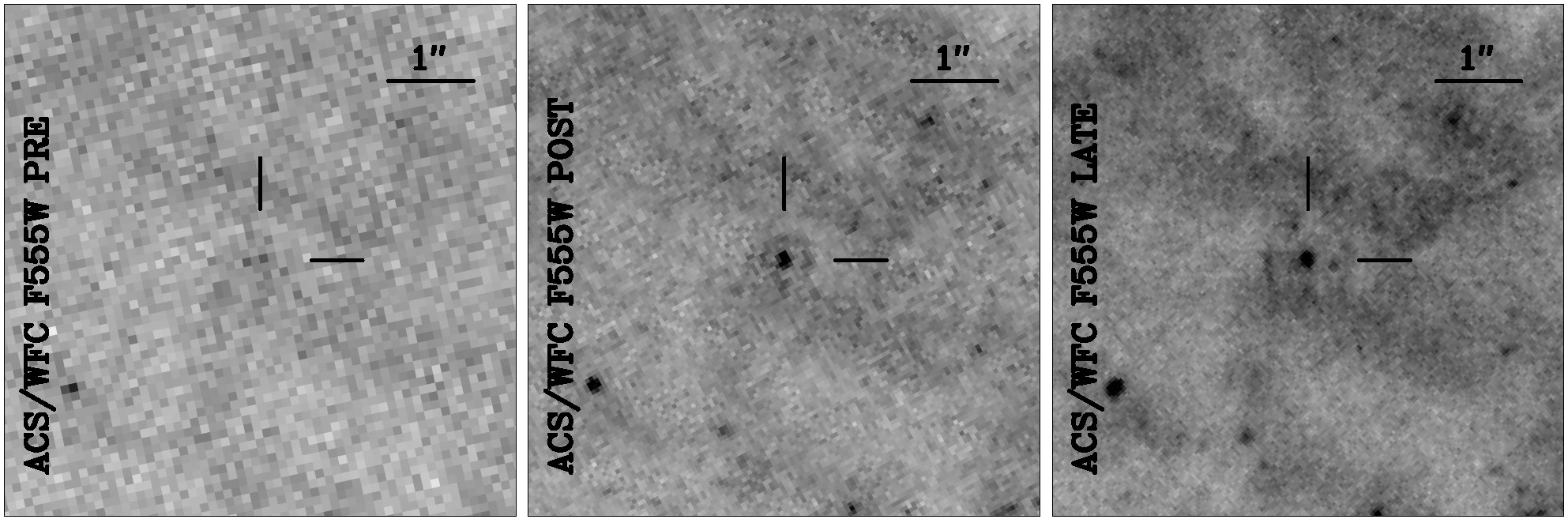}
\caption{HST imaging of the site of SN 1999ev.  From left to right:
  Pre-explosion WFPC2 WF2 $F555W$ image; Post-explosion ACS/WFC
  $F555W$ image; and late-time ACS/WFC $F555W$ image.}
\label{fig:obs:99ev_main}
\end{figure*}

\begin{figure}
\includegraphics[width=8cm]{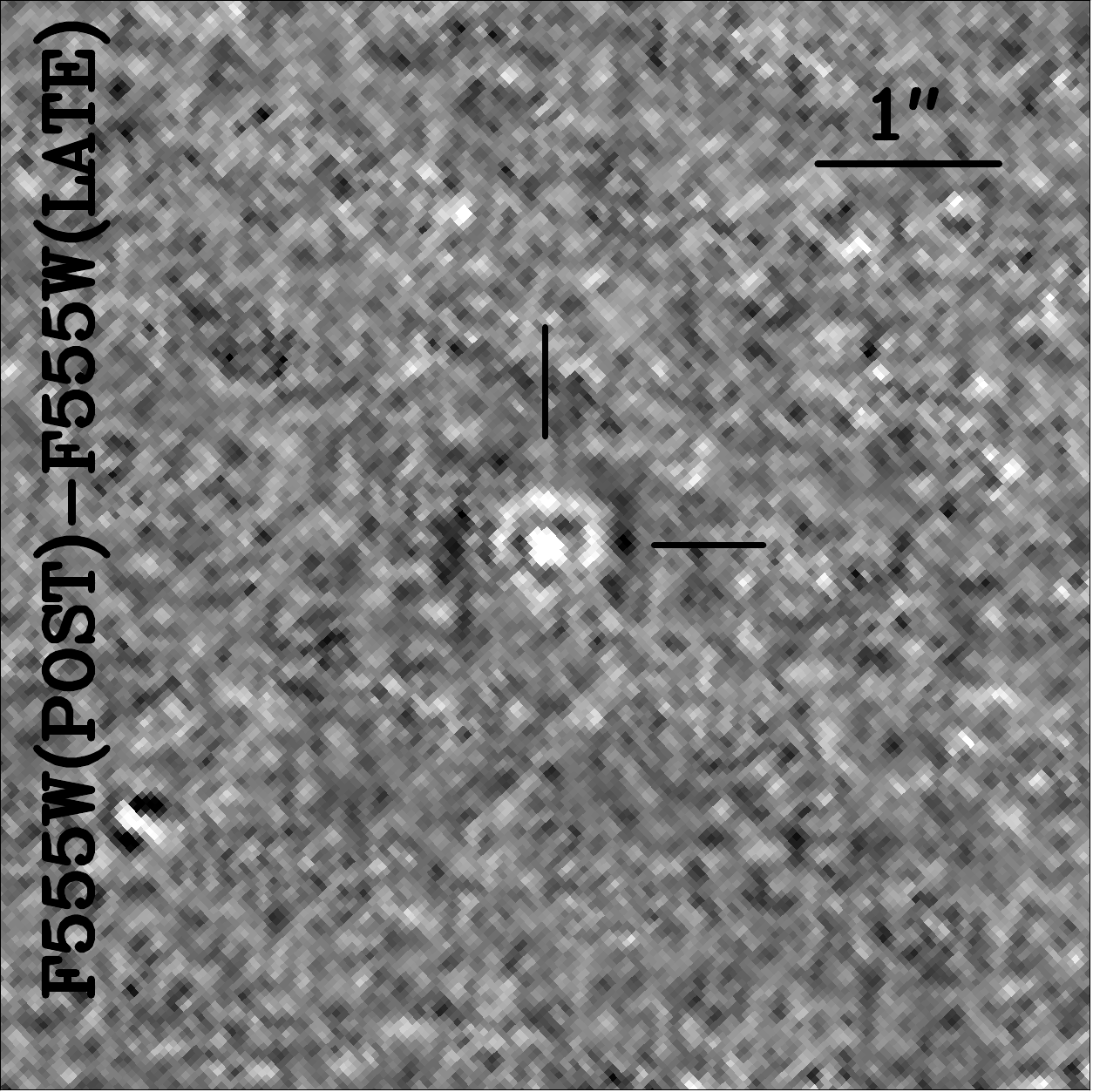}
\caption{Difference image between the post-explosion and late-time $F555W$ observations of the site of SN~1999ev.  The evolution of the expanding light echo is evident; the white echo appears in the post-explosion image, while the outer dark echo occurs in the late-time frame.}
\label{fig:obs:99ev_echo}
\end{figure}

Late-time ACS WFC $F435W$, $F555W$ and $F814W$ images (with pixel
scale $0.035\arcsec\,\mathrm{px^{-1}}$) of the site of SN~1999ev were acquired
on 2010 Nov 14 (11 years post-discovery).  A late-time image of the
site of SN~1999ev is shown as Fig. \ref{fig:obs:99evcol}.
A geometric transformation was calculated between the post-explosion
and late-time $F555W$ images using 24 commons stars, with an
uncertainty on the transformation of $\Delta r = 0.016\arcsec$.  A
source is recovered in the late-time images at the transformed
position of the SN as identified in the post-explosion images by
\citet{2005astro.ph..1323M}, as shown in Fig. \ref{fig:obs:99ev_main}.
The source is detected in all three filters: $m_{F435W}=25.63\pm0.07$,
$m_{F555W}=24.76\pm0.06$ and $m_{F814W}=23.65\pm0.05$.  The photometry
of the SN in the post-explosion images was recalculated, and the SN
was measured to have $m_{F435W}=24.79\pm0.10$,
$m_{F555W}=24.19\pm0.14$ and $m_{F814W}=23.49\pm0.10$.  We note that
the new measurement of the post-explosion $F555W$ photometry reported
here is slightly fainter than measured previously, although the
$F435W$ and $F814W$ magnitudes are approximately similar to those of
\citet{2005astro.ph..1323M}.  The $8.98$ years between the
post-explosion and late-time images reveals significant evolution in
the light echo discovered by \citet{2005astro.ph..1323M}, which has
expanded to a radius of $0.48\arcsec$ from $0.25\arcsec$ (as shown on
Figs. \ref{fig:obs:99ev_main} and \ref{fig:obs:99ev_echo}).

\begin{figure}
\includegraphics[width=8cm]{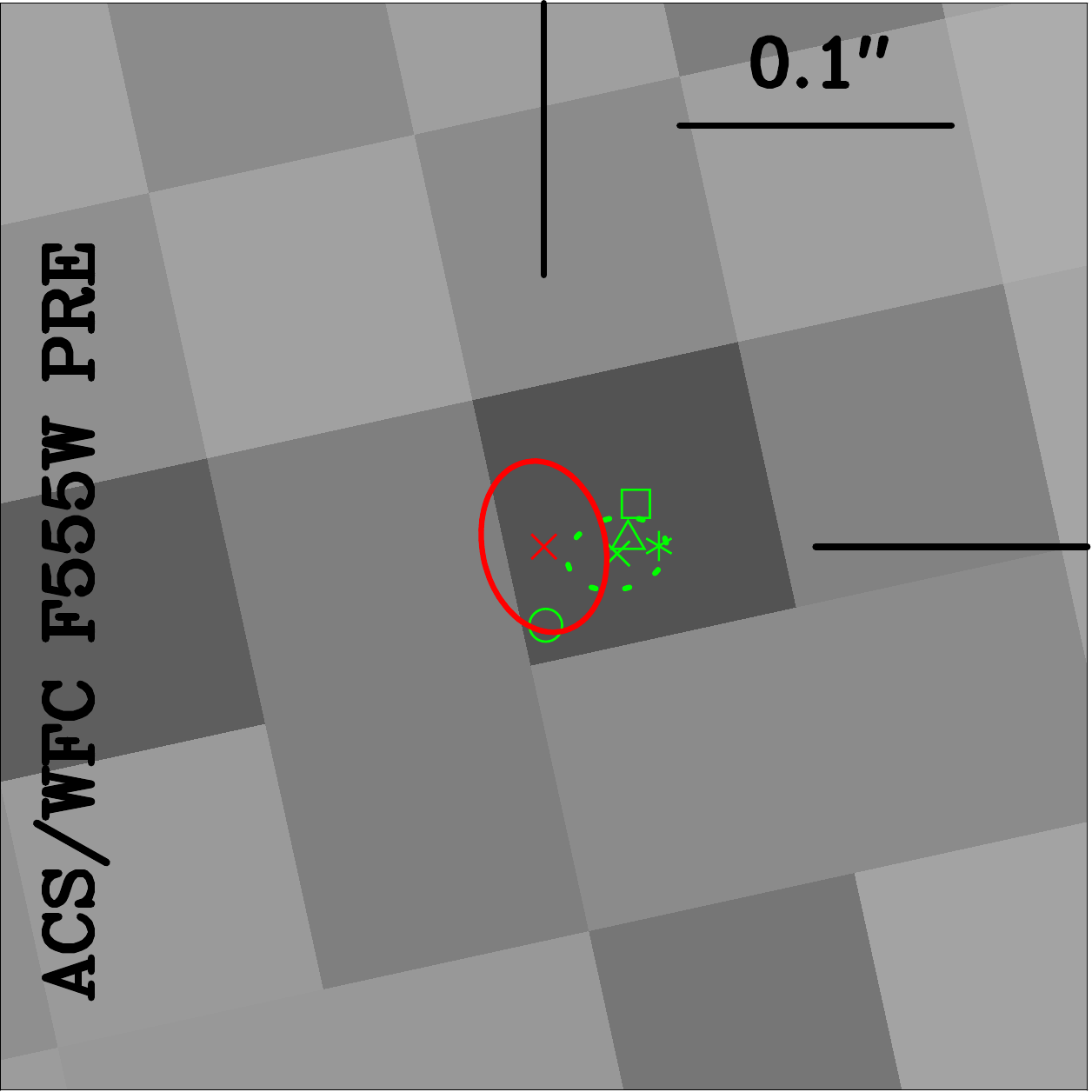}
\caption{Close-up of the pre-explosion site of SN 1999ev.  In red ($\times$) is the transformed position of the SN, derived from the post-explosion images, and the corresponding r.m.s. error ellipse.  Also shown (in green) are the four separate measures of the position of the pre-explosion source derived using the centroid ($\square$), Gaussian ($\triangle$) and optimal filter ($\circ$) centring algorithms in DAOphot and the position determined using HSTphot ($\ast$).  Also shown are the average ($\times$)  and the corresponding standard deviation ellipse of the four measurements.}
\label{fig:obs:99ev_zoom}
\end{figure}

A transformation was calculated between the post-explosion and
pre-explosion $F555W$ images using 24 stars (with a transformation
uncertainty of $\Delta r = 0.039\arcsec$).  In the pre-explosion
images, we identify the same source that \citet{2005astro.ph..1323M}
identified as the progenitor source with $m_{F555W}=24.66\pm0.17$.  In
addition, we also find a nearby source with $m_{F555W}=25.08\pm0.24$
located $0.2\arcsec$ (2 WF pixels) from the progenitor.  We note that,
for the period in which the pre-explosion observations were
conducted \footnote{http://www-int.stsci.edu/ftp/instrument\_news/WFPC2/Wfpc2\_hotpix/1995/\\ vary\_950113\_950211\_2.dat.Z},
the nearest logged warm pixel is 5 pixels away from the SN position
and not coincident with either the progenitor candidate or the nearby
object.  In the late-time image, however, we do not recover any source
at the corresponding transformed position.  The position of the
pre-explosion source at the SN position was estimated using the three
centring algorithms available to DAOphot (centroid, Gaussian
and optimal filter) and the position determined using HSTphot
PSF fitting.  The positions are shown, with respect to the transformed
position of the SN on the pre-explosion image, in
Fig. \ref{fig:obs:99ev_zoom}.  Although there is an apparent
discrepancy in the positions for the source and the transformed SN
position, the discrepancy is not significant.  It does, however, raise
concerns about how positions are determined on subsampled images such
as this pre-explosion WFPC2 WF2 $F555W$ image.  The position
determined using the optimal filter centring algorithm is noticeably
different from the other three positions derived from the
pre-explosion image and is offset in the direction of the nearby
apparent neighbouring star.  The standard deviation of the four
measurements made on the pre-explosion image is $0.022\arcsec$.  Given
the apparent brightness of the source, however, the astrometric
uncertainty for the position derived using HSTphot alone may be
as large as $0.04\arcsec$ or 0.4 WF pixels \citep{dolphhstphot}.

\subsection{SN~2003gd}
\label{sec:res:03gd}
SN~2003gd was discovered by R. Evans on 2003 Jun 12.82, in the galaxy
M74 \citep{2003IAUC.8150....2E}.  \citet{2003IAUC.8152....1K}
spectroscopically classified 2003gd as being a Type II SN,
approximately 2 days post-explosion.  Subsequent photometric and
spectroscopic observations of SN~2003gd, however, showed it to be a
Type IIP SN discovered at the end of the plateau phase
\citep{2005MNRAS.359..906H}.  \citet{2003IAUC.8152....4S} made a
preliminary identification of the progenitor in pre-explosion $HST$
WFPC2 $F606W$ and Gemini GMOS-N $i'$ images.  As such, SN~2003gd was
the third SN, after SNe~1987A and 1993J, to have a progenitor
identified in fortuitous pre-explosion images.  Independent analyses
by \citet{smartt03gd} and \citet{2003PASP..115.1289V} showed the
candidate progenitor to be a RSG, corresponding to a star
with initial mass of $\sim 8-9M_{\odot}$.  The confirmation of this
star as the progenitor was finally provided in 2009, when the star was
observed to no longer be present in late-time Gemini GMOS-N
$i^{\prime}$ images \citep{2009Sci...324..486M}; making it the first
conclusively confirmed RSG progenitor for a Type IIP SN.

\begin{figure}
\includegraphics[width=8cm]{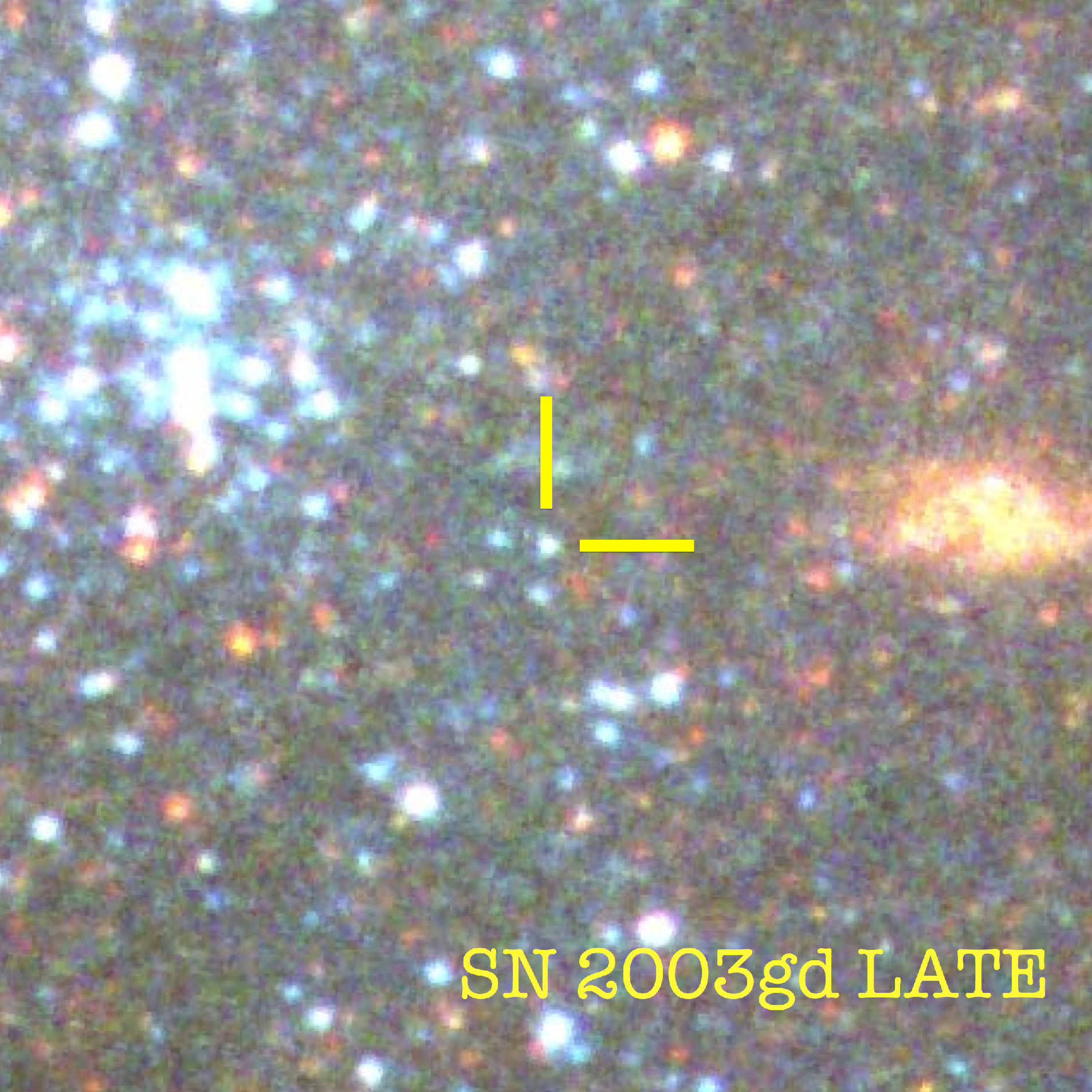}
\caption{Late-time colour image (composed of $F435W$, $F555W$ and $F814W$ images) of the area of M74 around the site of SN~2003gd.  The image has dimensions $8\arcsec \times 8\arcsec$, and is oriented such that North is up, and East is left.  The position of SN~2003gd is located at the centre of the image.}
\label{fig:obs:03gdcol}
\end{figure}

\begin{figure*}
\includegraphics{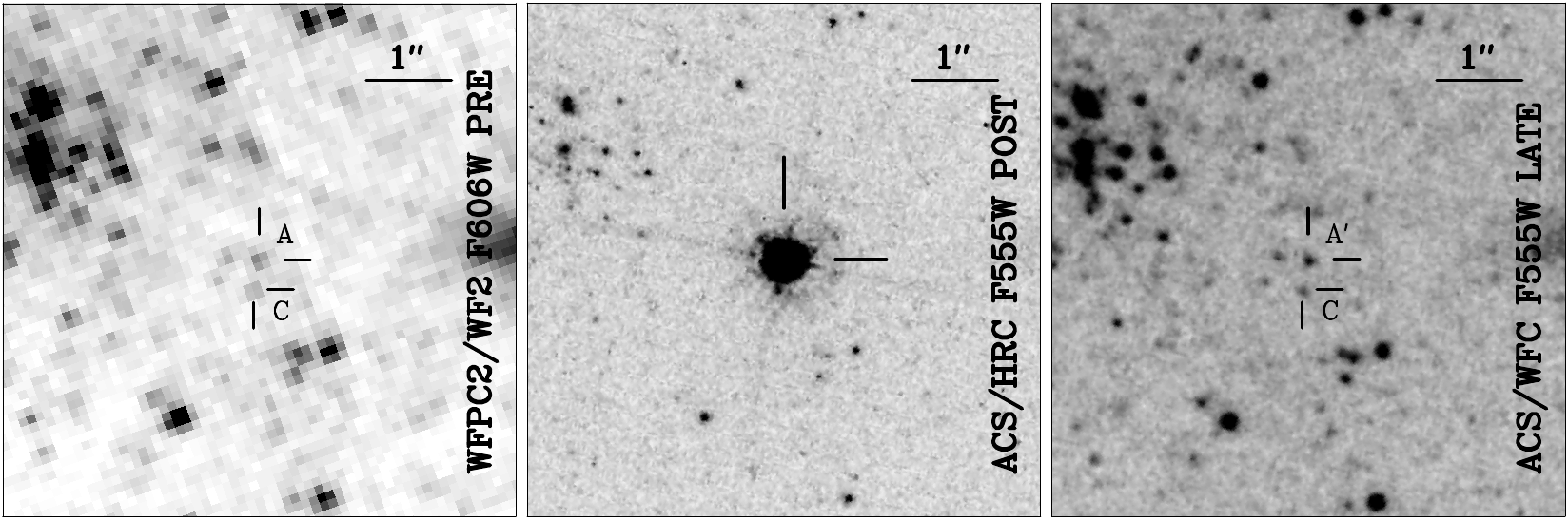}
\caption{HST imaging of the site of SN 2003gd.  From left to right: Pre-explosion WFPC2 WF2 F606W image; Post-explosion ACS/HRC F555W image; and late-time ACS/WFC F555W image.}
\label{fig:obs:03gd_main}
\end{figure*}

\begin{figure*}
\includegraphics{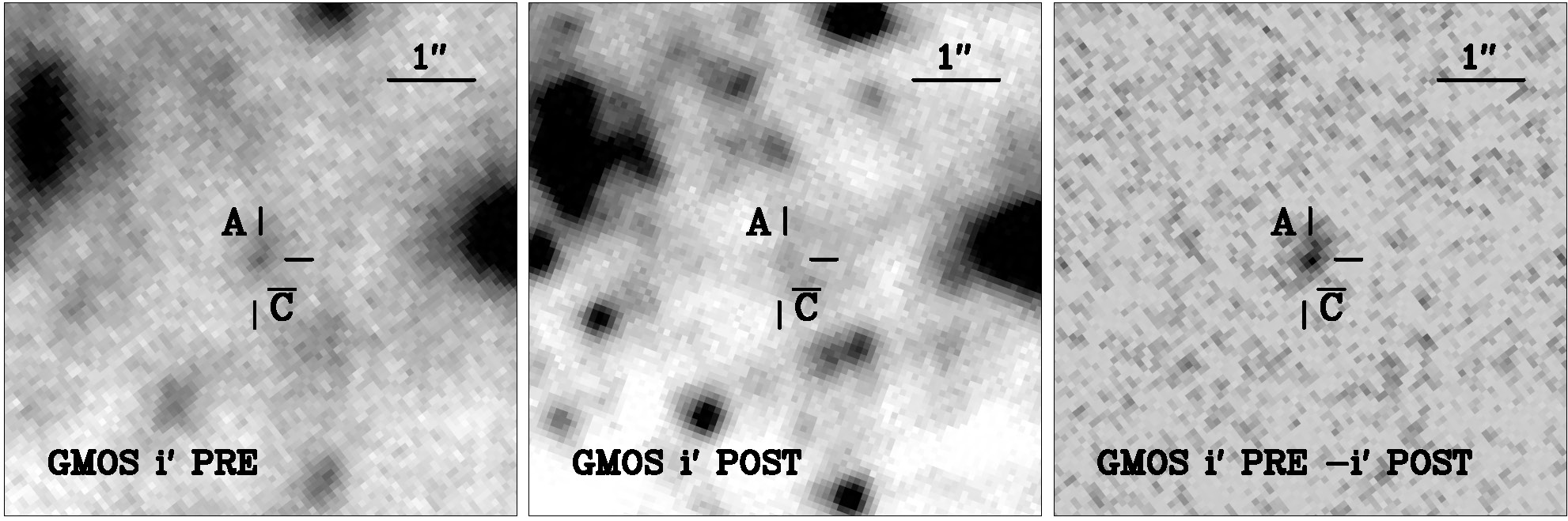}
\caption{Pre-explosion and late-time Gemini $i^{\prime}$ observations of the site of SN~2003gd (for further details see \citealt{2009Sci...324..486M}).}
\label{fig:obs:03gd_gmos}
\end{figure*}

Late-time ACS $WFC$ observations of the site of SN~2003gd were
acquired on 2010 Nov 14, and are presented on Figure
\ref{fig:obs:03gdcol}.  The SN position in the late-time images was
determined with respect to the ACS HRC post-explosion images, with an
uncertainty on the transformation of $0.010\arcsec$.  The position of
the SN on the pre-explosion HST and Gemini images, as presented by
\citet{smartt03gd} and \citet{2009Sci...324..486M}, were recalculated
to within $0.028\arcsec$ and $0.025\arcsec$ respectively.  The
pre-explosion, post-explosion and late-time data $F555W$ images are
shown on Figure \ref{fig:obs:03gd_main} and, for completeness, we also
show the corresponding $i^{\prime}$ data presented by
\citet{2009Sci...324..486M} as Figure \ref{fig:obs:03gd_gmos}.

In the late-time images we observe a source, termed Source
$A^{\prime}$, that is clearly recovered in all late-time HST images at
the transformed SN position, with magnitudes $25.93\pm0.04$,
$25.42\pm0.05$ and $24.90\pm0.04$ in the $F435W$, $F555W$ and $F814W$
filters respectively.  This source was detected in late-time Gemini
GMOS-N $g^{\prime}$ and $r^{\prime}$ images, but not recovered
significantly in the corresponding $i^{\prime}$ image, acquired on
2008 Sep 06.  \citet{2009Sci...324..486M} measured
$g^{\prime}=25.10\pm0.04$ and $r^{\prime}=24.49\pm0.05$ (in Vega
magnitudes) for the source at the SN position, and placed a detection
limit of $i^{\prime}>25.9$.  This source was also observed with
$HST$ WFPC2 on 2007 Aug 11 (for program $GO-11229$; PI: M. Meixner).
Photometry of these images using the HSTphot package yielded
$m_{F622W}=24.52\pm0.08$ and $m_{F814W}=25.22\pm0.26$.  We note that
this photometry is approximately $0.4$ magnitudes brighter than the
photometry of the same images reported by \citet{2012ApJ...744...26O}.
We confirmed our photometry using ISIS, determining the flux
difference between the source in the 2007 WFPC2 $F814W$ image and
our late-time ACS $F814W$ image is consistent with the photometry
conducted on the images directly.  
 
We recalculated the photometry of the source at the SN position in the
pre-explosion WFPC2 $F606W$ image, labeled Source $A$ by
\citet{smartt03gd}, using HSTphot finding
$m_{F606W}=25.06\pm0.06$.  \citet{2009Sci...324..486M} derived the
$i^{\prime}$ magnitude of the progenitor, using image subtraction
techniques (see Figure \ref{fig:obs:03gd_gmos}), of $23.85\pm0.04$
(with a possible 0.15 magnitude systematic uncertainty on underlying
residual flux in the late-time Gemini image).  Unlike the obviously
red Source $A$ observed in the pre-explosion $HST$ WFPC2 and Gemini
GMOS images, it is apparent that Source $A^{\prime}$ in the late-time
images is a blue-yellow object (see Figure \ref{fig:obs:03gdcol}).
Even taking into account a colour correction between the late-time
$F814W$ and Gemini GMOS $i^{\prime}$ photometry (see
Fig. \ref{fig:obs:colcor}), a significant increase in brightness is
evident between the two observations separated by two years.


\subsection{SN~2004A}
\label{sec:res:04A}
SN~2004A was discovered by K. Itagaki \citep{2004IAUC.8265....1N} on
2004 Jan 9.4 in the galaxy NGC 6207.  \citet{2004IAUC.8266....2K}
spectroscopically classified the SN as a being a young Type II SN.
\citet{2006MNRAS.369.1303H} presented photometric and spectroscopic
observations of SN~2004A and showed it to be consistent with other
normal Type IIP SNe, such as SN~1999em.
\citeauthor{2006MNRAS.369.1303H} also presented an analysis of the
pre-explosion $HST$ WFPC2 observations of the site of SN~2004A from
2001 Jul 02, in conjunction with post-explosion ACS WFC observations
of the SN acquired on 2004 Sep 23.  A source was barely recovered at
$4.7\sigma$ at the SN position in the pre-explosion $F814W$ image.
There was no corresponding source in the pre-explosion $F606W$ image,
consistent with a star with $F606W-F814W > 1.05$.
\citeauthor{2006MNRAS.369.1303H} concluded that if this was the
progenitor star an RSG with initial mass $9^{+3}_{-2}M_{\odot}$;
although given concerns about the significance of the detection of the
source at the SN position, \citeauthor{2006MNRAS.369.1303H} placed a
conservative limit on the initial mass of an undetected progenitor of
$<12M_{\odot}$.

\begin{figure*}
\includegraphics{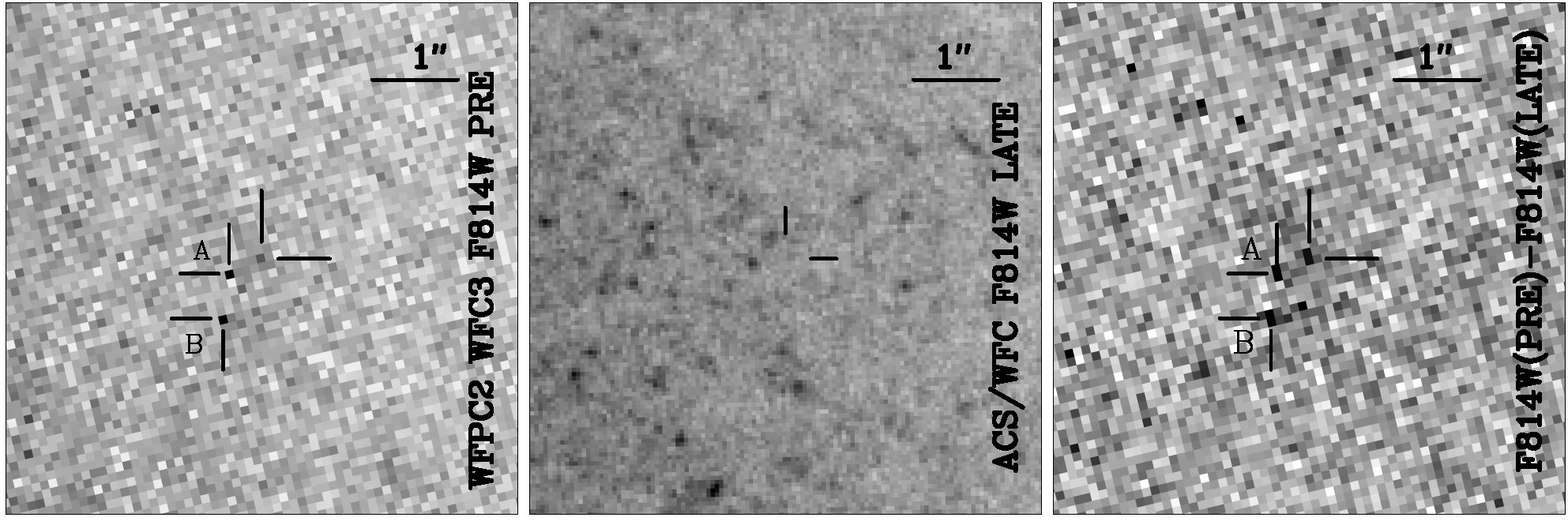}
\caption{The site of SN~2004A in NGC~6207.  Left) Pre-explosion WFPC2
  $F814W$ image from 2001 Jul 02 with scale
  $0.1\arcsec\,\mathrm{px^{-1}}$.  The centred cross hairs mark the
  transformed position of SN~2004A and the nearby sources $A$ and $B$
  are known hot pixels on the WFPC2 WF2 chip.  Centre) Late-time
  ACS/WFC $F814W$ image.  Right) Difference image between the late and
  pre-explosion $F814W$ images.  The observed residual sources
  correspond to the two hot pixels, $A$ and $B$, and the progenitor.}
\label{fig:obs:04Amain}
\end{figure*}

Late-time observations of the site of SN~2004A were acquired on 2010
Sep 09, approximately 6.7 years post-discovery.  The late-time $F814W$
observation and the corresponding pre-explosion $F814W$ observation
are presented on Figure \ref{fig:obs:04Amain}.  The pre-explosion and
late-time observations of the site of SN~2004A are presented in Figure
\ref{fig:obs:04Amain}.  Using the post-explosion $F814W$ image,
acquired on 2004 Sep 23 with ACS WFC, the position of the SN on the
pre-explosion and late-time frames was determined to within
$0.021\arcsec$ and $0.028\arcsec$, respectively.  The SN is not
detected significantly in any of the late-time ACS WFC images.  The
$3\sigma$ detection limits at the SN position, in the late-time
images, were evaluated with artificial star tests to be
$m_{F435W}=27.65$, $m_{F555W}=27.5$ and $m_{F814W}=26.85$ mags.  These
limits are consistent with the expected depth for these images
predicted by the ACS imaging exposure time
calculator \footnote{http://etc.stsci.edu/etc/input/acs/imaging/}.

In the pre-explosion $F814W$ image, HSTphot finds a source within
$0.034\arcsec$ of the transformed SN position with
$m_{F814W}=24.48\pm0.19$ mags, detected with a signal-to-noise ratio of
5.6 (as shown on Figure \ref{fig:obs:04Amain}).  The positional
uncertainty is slightly larger than the formal $1\sigma$ uncertainty
of the geometric transformation alone.  The ability of HSTphot
to determine the position of objects of such brightness, however, is
limited, such that the expected uncertainty on the position of the
object on the pre-explosion $F814W$ image is $\Delta r
\geq 0.060\arcsec$ \citep{dolphhstphot}.  We note that this is the
same object identified by \citet{2006MNRAS.369.1303H} as the possible
candidate progenitor, although they measured the source to be $\sim
0.2$ magnitudes brighter with a larger photometric uncertainty (we
note, also, they did not correct their reported photometry to an
infinite aperture, such that the discrepancy is a further $0.05$ magnitudes
larger).  We also find the two sources $A$ and $B$ identified by
\citet{2006MNRAS.369.1303H} are hot
pixels \footnote{http://www.stsci.edu/hst/wfpc2/analysis/wfpc2\_hotpix.html}.
The pixel
immediately adjacent to the pixel hosting the majority of the
progenitor candidate's flux is a warm pixel; however this pixel has a low dark
current (with low
variability) \footnote{http://www-int.stsci.edu/instruments/wfpc2/Wfpc2\_hotpix/2001/\\
vary\_010617\_010711\_3.dat.Z}
and was corrected by the OTFR pipeline.  We determined the $50\%$
completeness level for $3\sigma$ detections for the pre-explosion
images using artificial star tests conducting the HSTphot.
Artificial stars were placed in a 5px radius around the transformed SN
position.  We find the corresponding detection limits to be
$m_{F606W}=26.25\pm0.40$ and $m_{F814W}=25.10\pm0.45$ mags.

The late-time $F814W$ image was subtracted from the pre-explosion
image, and the difference images is presented on
Fig. \ref{fig:obs:04Amain}.  We find residuals at the location of the
progenitor source and the two hotpixels.  The disappearance of the
pre-explosion progenitor candidate in these late-time images confirms
the authenticity of the object as the progenitor.  We derived
photometry of the residual at the transformed SN position of
$m_{F814W}=24.57\pm0.12$ mags, including a systematic uncertainty of $\sim
10\%$.  Using the results from the artificial star tests, we derived a
CTI correction of $-0.216\pm0.013$, yielding a final magnitude for the
progenitor of $m_{F814W}=24.36\pm0.12$ mags.  This is similar to the
photometry derived by \citet{2006MNRAS.369.1303H}, although for very
different reasons and improved precision.

A similar difference image was determined for the pre-explosion
$F606W$ image and the late-time $F555W$ image.  No significant
residual was found in the difference image as expected, given the
absence of a source at the transformed SN position in the
pre-explosion image.  Due to differences between the pre-explosion $F606W$ and late-time $F555W$ filter transmission functions, we did not use the
difference image to derive detection limits for the progenitor.

\subsection{SN~2005cs}
\label{sec:res:05cs}
SN~2005cs was discovered by \citet{2005IAUC.8553....1K} on 2005 Jun
27.933 in the galaxy M51.  \citet{2005IAUC.8555....1M}
spectroscopically classified SN~2005cs as being a young Type II SN.
\citet{2005IAUC.8555....2R} provisionally identified a blue supergiant
in the field as a possible candidate for the progenitor, although
later analysis (in conjunction with high resolution post-explosion HST
ACS HRC images) by \citet{2005MNRAS.364L..33M} and
\citet{2005astro.ph..7394L} found the progenitor star to be a RSG
with initial mass $M_{ZAMS}\sim 8M_{\odot}$.

\begin{figure}
\includegraphics{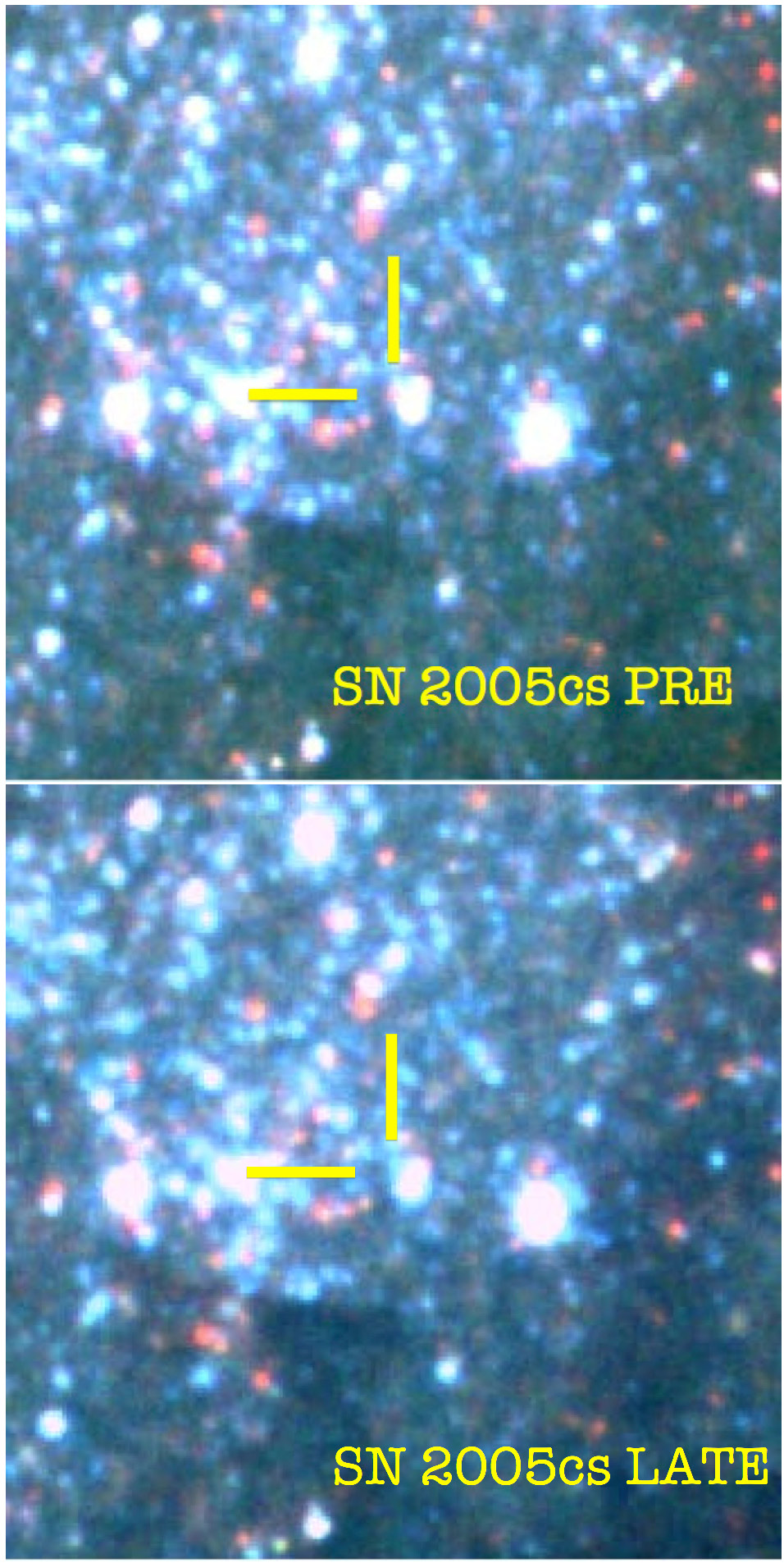}
\caption{Colour images of the site of SN 2005cs using ACS/WFC before explosion and at late-times.   The position of the SN is indicated by the cross-hairs.  In the pre-explosion image a red source is clearly visible at the SN position, and is found to be absent in the late-time image.  Each image has dimension $6\arcsec \times 6 \arcsec$, and is oriented such that North is up, East is left.}
\label{fig:obs:05csmain}
\end{figure}

\begin{figure*}
\includegraphics{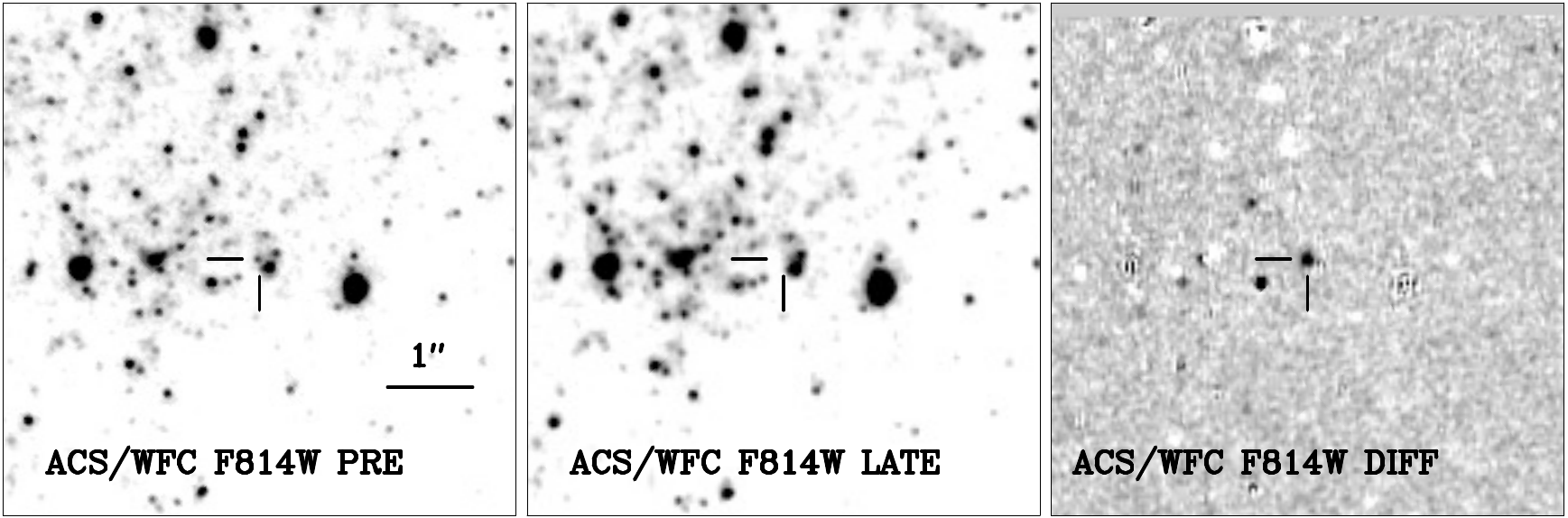}
\caption{ACS WFC $F814W$ observations of the site of SN 2005cs.  Left)
  Pre-explosion $F814W$ image (0.44 years prior to discovery).  The
  cross hairs indicate the transformed position of the SN and the
  identified progenitor candidate.  Centre) Late-time $F814W$ image
  (5.09 years post-discovery) with the progenitor candidate absent.
  Right) Difference image, between the pre-explosion and late-time
  $F814W$ observations, clearly showing a residual at the SN
  position.}
\label{fig:obs:05cs_panel}
\end{figure*}

Late-time observations of the site of SN~2005cs were acquired on 2010
Jul 30 with the ACS/WFC, 5.1 years post-discovery.  A comparison of
the pre-explosion and late-time observations of the site of SN~2005cs
is shown as Figs. \ref{fig:obs:05csmain} and \ref{fig:obs:05cs_panel}.
In this case, the late-time observations exactly match the
pre-explosion observations, using the same filters and detectors.  The
pre-explosion and late-time observations were drizzled to a final
common pixel scale of $0.035\arcsec$.  Utilising post-explosion ACS
HRC observations of SN~2005cs, the SN position was located on the
pre-explosion and late-time images to within $0.007\arcsec$ and
$0.004\arcsec$, respectively.

Direct photometry of the source detected at the SN position in the
pre-explosion $F814W$ yielded $m_{F814W}=23.382\pm0.048$, which is
$~0.1$ magnitude fainter than reported by \citet{2005MNRAS.364L..33M},
but $\sim 0.3$ magnitudes brighter than reported by
\citet{2006ApJ...641.1060L}.  In the late-time images, we do not
recover a source at the transformed SN position.  The detection limits
in these filters were probed using artificial star tests, yielding
$m_{F435W}=24.30\pm0.4$, $m_{F555W}=24.95\pm0.45$ and
$m_{F814W}=24.55\pm0.15$ mags.  These limits are particularly high,
relative to the expected depth for ACS/WFC images of these durations,
due to extended emission from the nearby cluster overlapping SN
position.

As noted by \citet{2005MNRAS.364L..33M} and
\citet{2006ApJ...641.1060L}, this underlying emission can complicate
the determination of the photometry of the progenitor from the
pre-explosion imaging alone.  This highlights the importance of using
image subtraction techniques to accurately derive the progenitor
photometry (by subtracting the background emission that is constant at
both epochs).  The difference image between the pre-explosion and
late-time $F814W$ observations is presented on
Fig. \ref{fig:obs:05cs_panel}.  We measure the brightness of the
progenitor to be $m_{F814W}=23.62\pm0.07$ mags, which is fainter than
the brightness determined from direct photometry of the pre-explosion
source (see above).  This magnitude is also significantly fainter than
the photometry of \citet{2005MNRAS.364L..33M}, and slightly brighter
than the photometry of \citet{2005astro.ph..7394L} (who attempted to
account for pre-explosion flux at the SN position due to the nearby
cluster).  We note that we find no significant source at the SN
position in the corresponding $F435W$ and $F555W$ difference images.

\begin{figure}
\includegraphics[width=6.5cm]{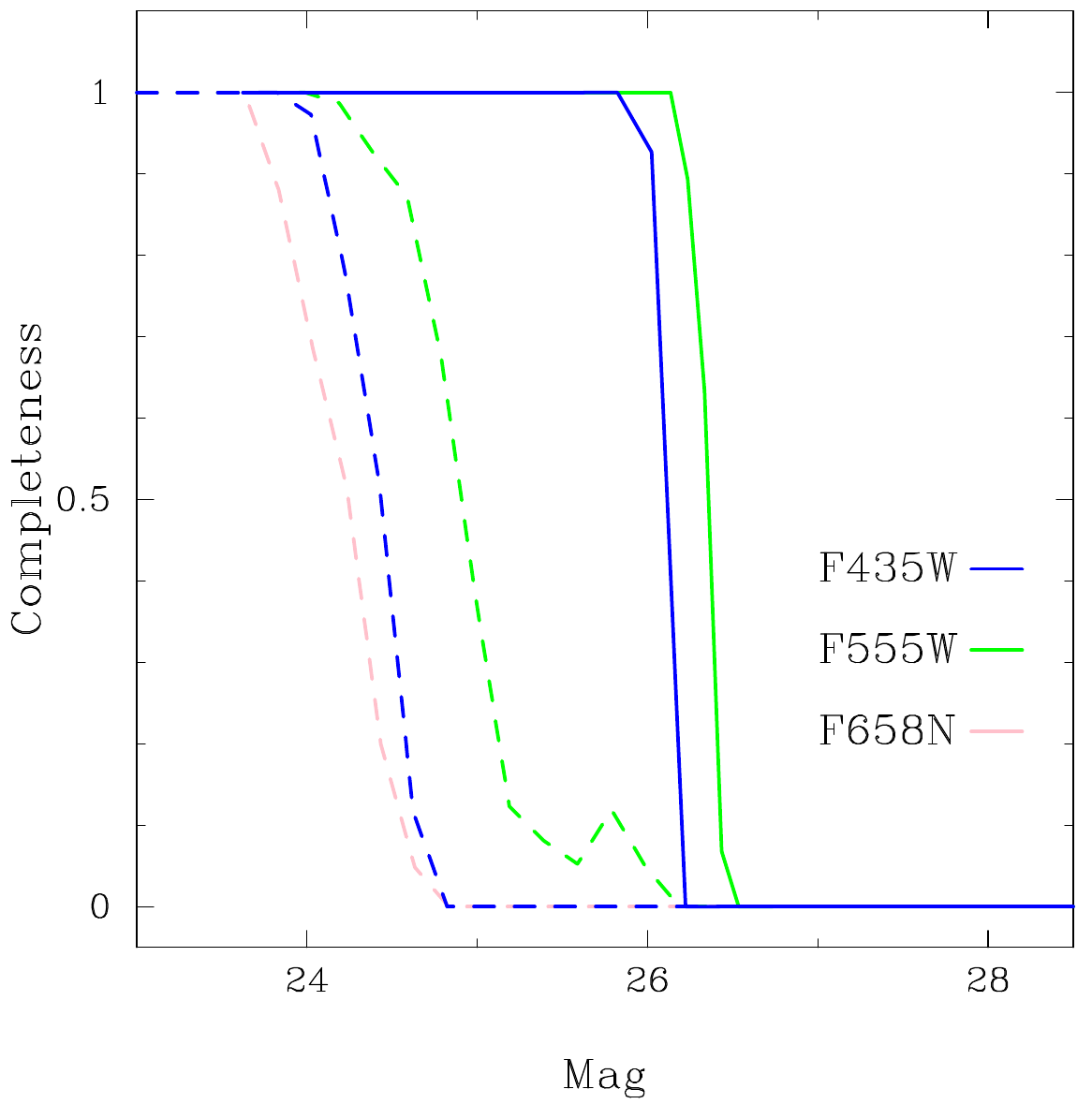}
\caption{Detection completeness functions for pre-explosion $ACS WFC$ observations of the site of SN~2005cs for the $F435W$, $F555W$ and $F658N$ filters (in which the progenitor was not detected).   The dashed curves indicate the completeness functions derived from attempted direct recovery of artificial stars on the original pre-explosion images, while solid curves are for detection limits derived in conjunction with image subtraction techniques.}
\label{fig:obs:2005cs_comp}
\end{figure}

Artificial star tests, in conjunction with image subtraction
techniques, were used to derive alternative detection limits (see
Section \ref{sec:obs:lim}) for the pre-explosion $F435W$ and $F555W$
images.  In the absence of a corresponding late-time $F658N$ ACS/WFC
image, the detection limit on the pre-explosion $F658N$ image could
only be derived using direct recovery of artificial stars on the
pre-explosion frame.  The photometric completeness functions for the
pre-explosion observations in which the progenitor was not detected is
shown on Fig. \ref{fig:obs:2005cs_comp} and presented in Table
\ref{tab:obs:2005cs_comp}.  The detection of a residual in {\sc isis}
difference images requires only a significant degree of residual flux
at the SN position and is less dependent on the amount of background
flux than the direct recovery of artificial stars.  There are
differences between the detection limits derived on the pre-explosion
images here and the limits presented by \citet{2005MNRAS.364L..33M}
and \citet{2006ApJ...641.1060L}.  These studies used combinations of
the analytical noise expression and artificial star tests, and treated
the effect of flux from the nearby cluster differently.  We note that
our detection limits derived using image subtraction techniques are
significantly deeper, highlighting the importance of late-time images
even in cases where detections of the progenitor are dubious or
unavailable.

We also find that there are a number of other sources that are clearly
variable between the pre-explosion and late-time images in the
vicinity of SN~2005cs (as shown on Fig. \ref{fig:obs:05cs_panel}).
Inspection of the pre-explosion and late-time images shows that the
apparent residuals in the difference images are associated with stars
which are clearly brighter or fainter in the late-time images compared
with the pre-explosion images.  Given the density of stars in this
field, compared with the sites of the other SNe considered here, it is
to be expected that there would be other variable sources in the field
around SN~2005cs.  The other residuals in the difference images are,
therefore, consistent with other real variables, but the progenitor
object is the only star to be absent in one of the two sets of the
images.


\begin{table}
\caption{\label{tab:obs:2005cs_comp} Detection limits for
  pre-explosion ACS/WFC $F435W$, $F555W$ and $F658N$ images of the
  site of SN~2005cs measured using artificial star tests and direct
  recovery and image subtraction techniques.  Previously derived
  detections limits from \citet{2005MNRAS.364L..33M} and
  \citet{2006ApJ...641.1060L} are shown for comparison}
\begin{tabular}{ccccc}
\hline\hline
       &    Recovery     &      ISIS    & \citeauthor{2005MNRAS.364L..33M} & \citeauthor{2006ApJ...641.1060L}\\
\hline
F435W  &  $24.35\pm0.20$ & $25.8\pm0.1$ &24.8 & $\cdots$\\
F555W  &  $24.85\pm0.30$ & $26.4\pm0.1$ &25   & 25.6    \\
F658N  &  $24.15\pm0.30$ & $\cdots$     &     & $\cdots$\\
\hline\hline
\end{tabular}
\end{table}

\subsection{SN~2006my}
\label{sec:res:06my}
SN~2006my was discovered on 2006 Nov 8.82UT by K. Itagaki
\citep{2006CBET..727....1N} in the galaxy NGC 4651.
\citet{2006CBET..737....1S} spectroscopically classified the SN as
being a Type II SN similar to SN~1999em. \citet{2007ApJ...661.1013L},
\citet{2008PASP..120.1259L} and \citet{2011MNRAS.410.2767C} analysed
the pre-explosion WFPC2 images of the site of SN~2006my.  All three
studies commented on the significant offset between the transformed SN
position and a nearby source recovered in the pre-explosion $F814W$
image.  \citeauthor{2008PASP..120.1259L} and
\citeauthor{2011MNRAS.410.2767C} concluded that the $F814W$ source was
unrelated to the SN, and that the progenitor was not detected in
either the pre-explosion $F555W$ or $F814W$ images.

\begin{figure*} 
\includegraphics{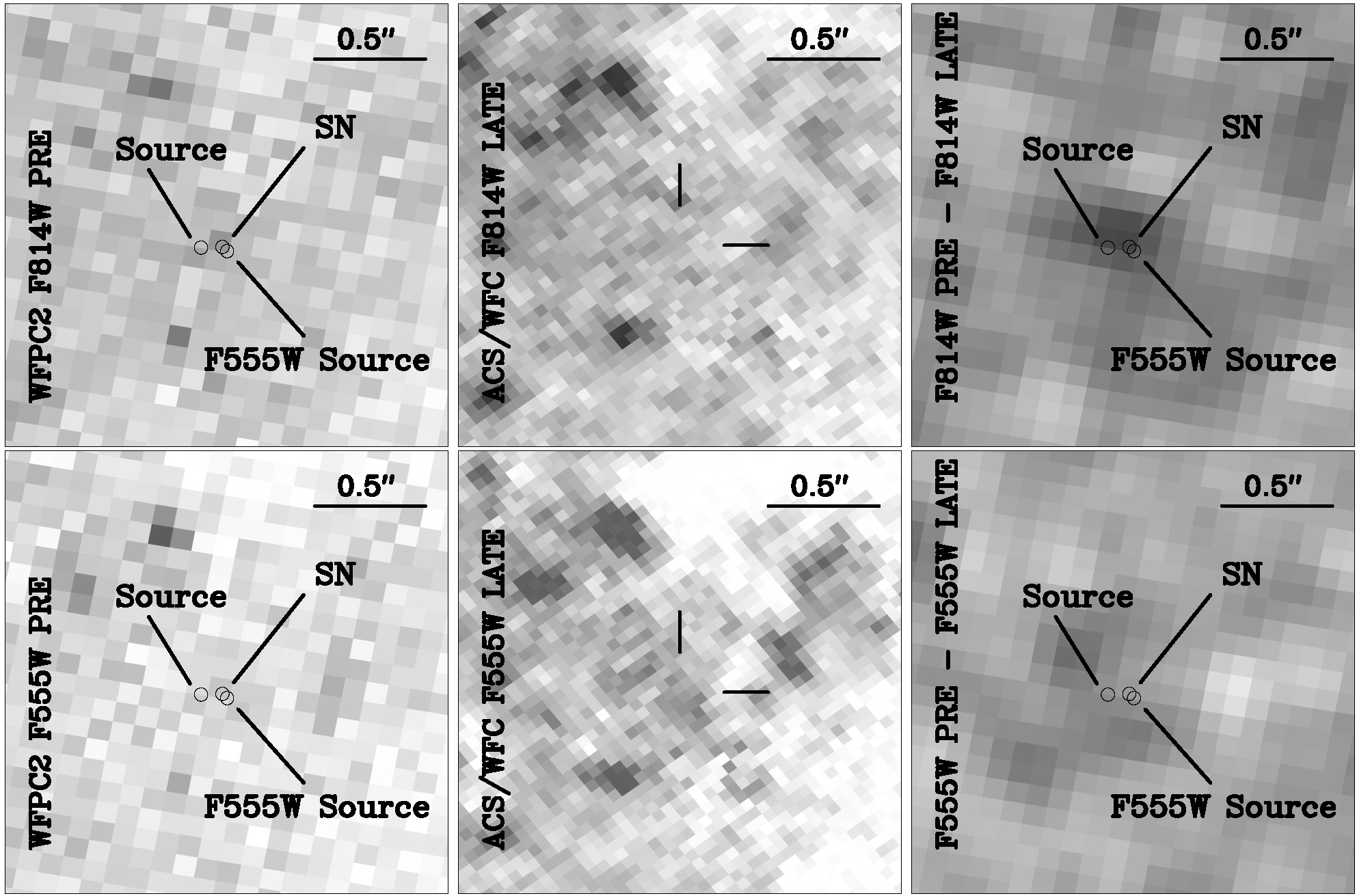}
\caption{Pre-explosion and late-time HST images of the site of SN 2006my in the $F814W$ (top row) and $F555W$ (bottom row) bands.  The transformed position of the SN and the positions of the sources detected in the pre-explosion $F555W$ and $F814W$ (labeled ``source'') images are indicated by the circles.}
\label{fig:res:06my}
\end{figure*}

The pre-, post-explosion and late-time $F555W$ and $F814W$ imaging of
the site of SN~2006my is shown on Fig. \ref{fig:res:06my}.  Following
the analyses presented by \citet{2007ApJ...661.1013L},
\citet{2008PASP..120.1259L} and \citet{2011MNRAS.410.2767C}, we
analysed the pre-explosion WFPC2 WF2 $F555W$ and $F814W$ covering the
position of SN 2006my.  The SN position, derived from post-explosion
WFPC2 PC1 images, was determined on the pre-explosion images using 19
common stars with a resulting uncertainty of $0.024\arcsec$; larger
than achieved by \citet{2011MNRAS.410.2767C}.  The transformed
position is found to be in the proximity of a cluster of bright pixels
and, as previously found by \citet{2007ApJ...661.1013L},
\citet{2008PASP..120.1259L} and \citet{2011MNRAS.410.2767C}, the
transformed position is not consistent with the position of the
nearest source found by HSTphot in the pre-explosion $F814W$
image (see Fig. \ref{fig:res:06my}).  The source detected by
  HSTphot is located $0.97$ pixels from the transformed SN position,
an offset significantly larger than the transformation uncertainty.
The transformed SN position is close to the position of a source in
the pre-explosion $F555W$ image but, as noted by the previous studies,
based on the sharpness value derived by HSTphot, it is not
consistent with a point source and is located $1.19$ pixels from the
source detected in the $F814W$ image.  The shift between the
pre-explosion $F555W$ and $F814W$ image was found to be very small:
$\Delta x = -0.0039$ and $\Delta y=0.0395$.

Using HSTphot, the pre-explosion $F814W$ source was measured to
have brightness $24.24\pm0.18$, which is approximately $0.2$
magnitudes brighter than found in the previous studies (which is due
to the use of different HSTphot settings used here).  Sections
of the late-time $F555W$ and $F814W$ images were transformed to match
the pre-explosion images; resampling the ACS WFC $0.05\arcsec$ pixels
to $0.1\arcsec$.  These images were processed using {\sc isis}, and
the resulting difference images are shown in Fig. \ref{fig:res:06my}.
{\sc isis} significantly detects a residual, at a distance of only
$0.11$ pixels from the transformed position of the SN.  Using 5
reference stars, and the DOLPHOT photometry derived for these
stars in the late-time $F814W$ image, we derive a magnitude of
$24.88\pm0.13$ mags for the residual in the difference image.
Artificial star tests on the pre-explosion images were used to derive
the CTI correction, for a star at the position of the observed
residual on the pre-explosion frame, of $0.016\pm 0.003$.  The
residual in the $F814W$ difference image corresponds, therefore, to an
object with $m_{F814W}=24.86\pm0.13$.  Given the coincidence of
residual with the transformed SN position, we conclude that the
progenitor was detected in the pre-explosion $F814W$ image.  The
fainter $F814W$ magnitude derived using ISIS, compared to our own HSTphot photometry of the pre-explosion images and the previously
reported values, and the apparent discrepancy in the position of the
pre-explosion source and the SN, most likely reflects that the source
in the pre-explosion $F814W$ image is a blend of the progenitor with a
source due East of the SN position (which also skews the apparent
position of the progenitor source in that direction).  In the analysis
of the pre-explosion and late-time $F555W$ images no residual was
found in the difference image.  The possible nature of the
pre-explosion $F555W$ source is revealed in the late-time images as a
complicated, extended background feature.

We used artificial star tests to probe the detection limit of the
pre-explosion $F555W$ and $F814W$ images within a 10 pixel radius of
the SN position; deriving $50\%$ completeness limits at $3\sigma$ of
$m_{F555W}=26.15\pm0.65$ and $m_{F814W}=25.05\pm0.75$ mags.

\section{Analysis}
\label{sec:ana}
In previous studies
\citep[e.g.][]{2005astro.ph..1323M,smartt03gd,2006MNRAS.369.1303H,2005astro.ph..7394L,2005MNRAS.364L..33M},
the photometric properties of the progenitor and surrounding stars
were derived through comparison with the ideal supergiant colour
sequence presented by \citet{drill00}.  As noted by Maund (2013, in
prep.) there are significant deficiencies with this approach; such as
the requirement for colour transformation equations to transform the
observed photometry to the photometric system of \citeauthor{drill00}.
More recent studies
\citep[e.g.][]{2012AJ....143...19V,2011arXiv1106.2565M,2011arXiv1106.2897V,2012arXiv1204.1523F,2013arXiv1302.0170M}
have shown the benefit in fitting directly to SEDs constructed from
synthetic spectra with known parameters using the same filters as the
observations.

In considering the observed photometry of objects identified at the
position of the target SN in the pre-explosion images we utilised the
BIX nested sampling and BASIE Markov Chain Monte Carlo SED fitting
packages described by Maund (2013, in prep.).  These two packages
allow us to comprehensively probe the effects of different stellar
parameters on the interpretation of the observed photometry on a
densely sampled grid of model stellar photometry, in the native filter
system of the observations.  By design, both of these SED fitting
packages can handle detections and upper limits simultaneously,
although for limited data (i.e. the number of detections is less than
the number of free parameters) only the BASIE code can be used to
explore the allowed parameter space rather than locate a unique
solution.  Crucially, we can explore the degeneracies between the
parameters (such as temperature and reddening), and implicitly account
for correlations between the temperature and bolometric luminosity
(through the bolometric correction).

Here we use two families of stellar SED models: the
ATLAS9 \citep{2004astro.ph..5087C} and MARCS \citep{marcsref} models.
Synthetic photometry of these models was conducted using our own
codes.

As we expect the progenitors to be cool RSGs ($<4500K$), we interpret
the observed photometry (and upper limits) with respect to the
$5M_{\odot}$ spherical MARCS SEDs, which have been successfully
compared with observations of RSGs in a number of previous studies
\citep[e.g. see][]{2005ApJ...628..973L,2013arXiv1302.2674D}.  We
assume that RSGs are well described by models with surface gravity
$\log g = 0.0$, and fit for the effective temperature ($T_{eff}$) and
the reddening.  We consider the effects of foreground and host reddening to be
due to Galactic-like dust (parameterised by $E(B-V)$;
\citealt{ccm89}).  To constrain the effect of reddening due to dust
expected to be found around RSGs, we consider the reddening laws for
graphite or silicate dust (parameterised by the optical depth $\tau_{\nu}$), contained
in spherical shells with ratios for the inner and outer radii of
$R_{out}/R_{in}=2$ or 10, following \citet{2012ApJ...759...20K}.  For
each SN site we adopt models with metallicities appropriate for that
site (see Table \ref{tab:obs:sne}).

To provide an additional handle on the interstellar reddening towards
each progenitor, we also consider the reddening towards the stars
immediately surrounding the SN position.  Due to the distances of the
host galaxies, the surrounding stars, which are selected based on the
condition that they are detected in all three of the late-time images
for each SN, are expected to be luminous OB stars. For each of the
stars, the parameters $T_{eff}$ and $E(B-V)$ are derived with respect
to the ATLAS9 models.  We selected those models from the ATLAS9 grid
that are consistent with supergiant surface gravities
\citep{synphot}\footnote{http://www.stsci.edu/hst/HST$\_$overview/\\documents/synphot/AppA$\_$Catalogs4.html}.
Total reddenings are derived with respect to a \citet{ccm89}
$R_{V}=3.1$ reddening law, appropriate for reddening and extinction
due to interstellar dust.  The SED fits were conducted with the BIX
Nested Sampling package (as all stars, by selection, had three colour
photometry), such that the Bayesian evidence could be used to exclude
those stars with colours clearly inconsistent with OB stars.
Furthermore, to consider the SEDs of compact clusters we adopt the
model spectra produced using the {\sc starburst99} code
\citep{1999ApJS..123....3L}.

For stellar progenitors, we derive masses using our various luminosity
estimates (depending on the type of dust and star) following the technique of \citet{2008arXiv0809.0403S}.
\citeauthor{2008arXiv0809.0403S} use predicted luminosities for the
end-phases of STARS stellar evolution models \citep{eld04} to derive
initial masses for progenitors with luminosities constraints derived from the observations.  For a given luminosity, the progenitor is considered to
lie in the mass range bounded at one end by the most massive star to
end core He burning at that luminosity, and at the other by the least
massive star to proceed to model termination (the onset of core Ne
burning) at that luminosity \citep[see][and their
  Fig. 1]{2008arXiv0809.0403S}.  We use STARS models calculated at
integer initial masses, with the appropriate metallicities, and
interpolate to determine the luminosities at the end of core He
burning and the beginning of core Ne burning.  This scheme
characterises possible RSG progenitors.  We also note that,
at lower masses, some stars will undergo second dredge-up, causing
them to become Asymptotic Giant Branch (AGB) stars that are cooler but
more luminous than the similar mass stars that die as RSGs.
Based on the observed temperature range of RSGs, derived
using MARCS spectra \citep{2005ApJ...628..973L}, we use a temperature
threshold of $3400K$, above and below which we consider stars to be
RSGs or AGB stars, respectively.

In deriving posterior probability density functions (pdfs) for the
initial masses for the progenitors, we also consider the effect of
prior information from the IMF.  We apply a weighting factor to the
posterior pdfs $\propto M^{-2.35}$ for a \citet{1955ApJ...121..161S}
IMF, to follow the weighting scheme applied by
\citet{2008arXiv0809.0403S} for their analysis of the Type IIP SN
progenitor population.


\subsection{SN~1999ev}
\label{sec:ana:99ev}
\begin{figure}
\includegraphics[width=6.5cm]{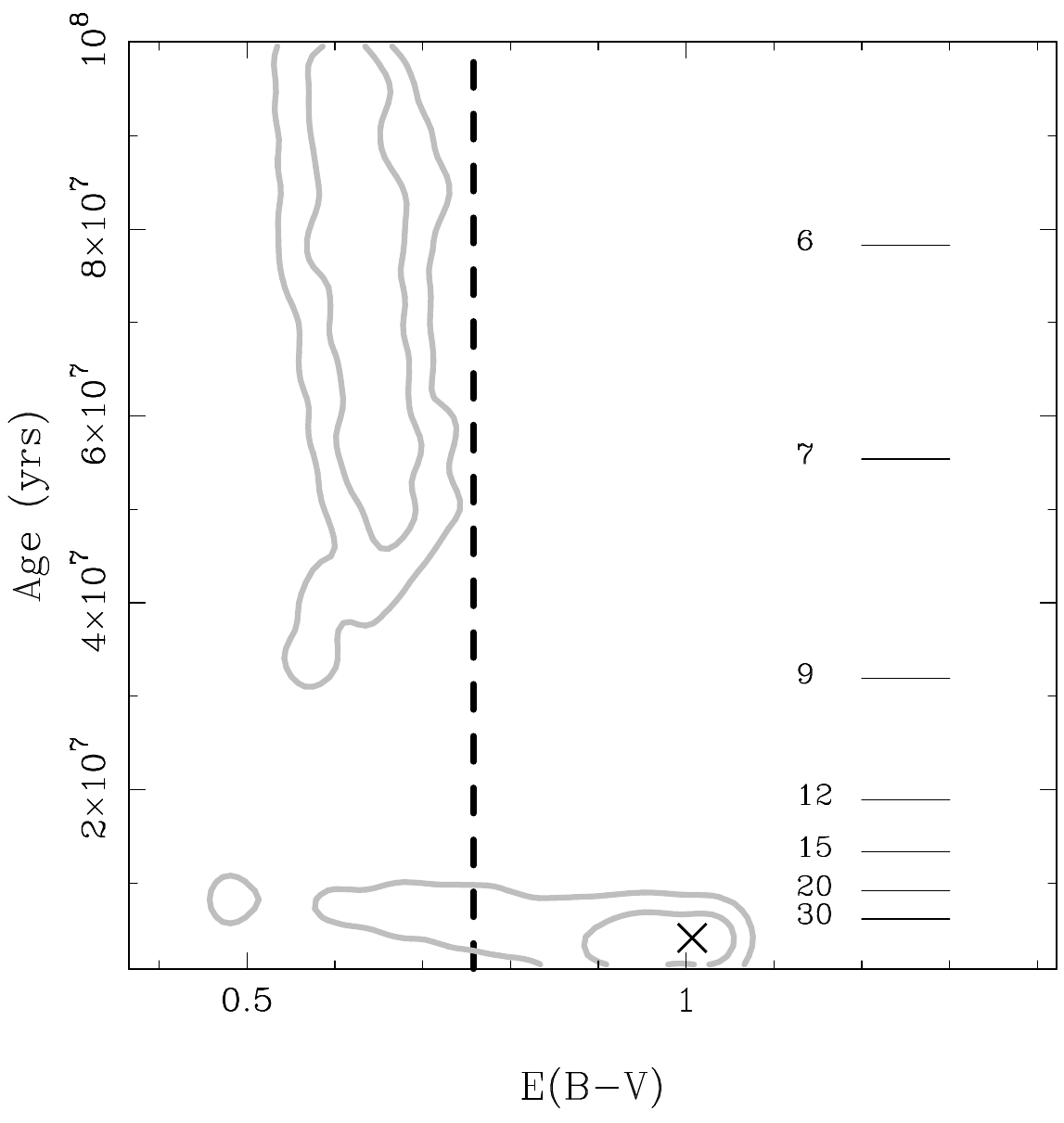}
\caption{The posterior probability distribution for SED fits of the
  late-time photometry of the source at the position of SN~1999ev to
  cluster SED models produced using {\sc Starburst99}.  The contours
  contain the $68\%$ and $95\%$ of the probability, while the most
  likely solution is indicated with the cross.  The vertical dashed
  line corresponds to $E(B-V)=0.76$; for $E(B-V) \geq 0.76$, the
  absolute magnitude of the source is $M_{F555W} < -8.6$.  The
  horizontal lines indicate the lifetimes of stars with a given
  initial mass as predicted by the Padova stellar evolution code
  \citep{2002A&A...391..195G}.}
\label{fig:obs:99ev_clus}
\end{figure}

The presence of a source in the late-time images precludes the use of
image subtraction techniques to further analyse the nature of the
pre-explosion source.  We note that the pre-explosion $F555W$
magnitude of the source at the SN position is of similar magnitude to
the source in the late-time $F555W$ image.  The difference between the
pre-explosion and late-time photometry is not significant with a
p-value of 0.71 (using a simple $z$-test).  This lends support to the
hypothesis that the source observed in the pre-explosion images at the
SN position has been recovered in the late-time images, and that the
original identification of the progenitor presented by
\citet{2005astro.ph..1323M} is, at least partially, incorrect.  We
suggest three possible scenarios for the nature of the source at the
SN position:
\begin{enumerate}
\item{The source at the SN position in the pre-explosion and late-time
  images is a host cluster that contained the now absent progenitor.}
\item{The source observed in the pre-explosion and late-time image is
  an unrelated star that is coincident with the line-of-sight to the
  SN.  While the late-time observations do suggest a large, young
  stellar population hidden by the large dust sheet, the
  determination of the likelihood of a chance alignment is non-trivial.
  Given the astrometric coincidence $<0.04\arcsec$, it is likely to be
  very low.}
\item{The source observed in the late-time images is an unresolved
  light echo, and the pre-explosion source has now disappeared.  Given
  the observation of evolving light echoes around the position of the
  SN, and the obvious amount of dust in the vicinity of the SN, the
  apparent late-time brightness may be due to a light echo from dust
  immediately behind the SN.  We find this scenario unlikely, as it
  requires the progenitor and light echo to have coincidentally
  similar brightness.}
\end{enumerate}
Given the nature of the late-time three-colour imaging it is not
possible to unambiguously distinguish between the different scenarios.
The late-time images were used to examine the consequences of the
progenitor residing in a host cluster that was observed in the
pre-explosion and late-time images.  The shape of the source in the
late-time images was measured using the {\sc ishape} package
\citep{1999A&AS..139..393L}.  In each filter band, {\sc ishape}
returned a significantly better fit with a Moffat function over a
delta function ($\chi^{2}(\mathrm{Moffat})/\chi^{2}(\mathrm{delta}) <
0.95$) and an effective radius $R_{eff} > 0.1\times$ the Full Width at
Half Maximum \citep{1999A&AS..139..393L}.  The effective radius of the
source was measured to be $0.99^{+0.3}_{-0.55}$,
$1.82^{+0.56}_{-1.07}$ and $1.18^{+0.12}_{-0.57}$ pixels in the
$F435W$, $F555W$ and $F814W$, respectively (with a pixel scale
corresponding to $2.7 \mathrm{pc\, px^{-1}}$ at the distance of NGC
4274).  The large error bars are symptomatic of the complexity of the
region hosting the SN, in particular with the proximity of light
echoes and the apparent faintness of the source.  The {\sc ishape}
analysis was also conducted on six nearby objects that were all found
to be consistent with point-like, stellar sources - suggesting that
{\sc ishape} does have the capability, under the conditions of the
late-time images, to differentiate extended sources from point-like
sources.  To further explore the implications of a host cluster for
the progenitor, the late-time $ACS$ photometry was compared with {\sc
  Starburst99} models \citep{1999ApJS..123....3L}; and the results of
this fit is shown as Figure \ref{fig:obs:99ev_clus}.  Given the three
colour photometry, there are two allowed solutions: a moderately
reddened older solution ($40-100\,\mathrm{Myr}$) implying $M_{ZAMS} <
9M_{\odot}$; and a heavily reddened younger solution
($<10\,\mathrm{Myr}$) implying $M_{ZAMS} > 20M_{\odot}$\footnote{We
  note that the {\sc starburst99} code uses the Padova stellar
  evolution \citep{2002A&A...391..195G} models, and the masses we
  report for the derived ages are from these models.}.  In addition,
given the criterion presented by \citet{2005A&A...443...79B}, that
point-like objects with $M_{V} < -8.6$ are more likely to be clusters
than individual bright stars, requires $E(B-V)\geq0.76$ (assuming an
$R_{V}=3.1$ Galactic reddening law).  In terms of shape, absolute
brightness and colours, the late-time source is consistent with a
cluster.  Using the photometry of nearby point-like sources, a
weighted-average reddening of $E(B-V)=0.95\pm0.32$ was measured.  The
large error bar is consistent with both the poor photometric errors
for each of the six nearby sources and the large scatter in reddenings
in the sample.  The reddenings towards these objects are significantly
greater than expected for just pure foreground Galactic reddening
($E(B-V)=0.2$) towards NGC~4274.  This may reflect a complex dust
distribution where some of the sources are in front of the dust sheet
and others may be embedded.  This suggests that there might be
significant reddening towards the source at the SN position, however
this is not conclusive.

If the late-time source is, in fact, a light echo, then the late-time
images do not provide any further insight into the properties of the
source in the pre-explosion images.  Our inability to the confirm the
disappearance of the progenitor also means that the late-time images
cannot be used to rule out the possibility that the object in the
pre-explosion and late-time images is an unrelated object in the
line-of-sight.  High-resolution near-infrared observations, with the
HST, could be used to probe the nature of the
stellar population behind the dust sheet (to examine the density of
objects along the line-of-sight) as well as provide further
constraints on the nature of the source in late-time images and nature
of the obscuring dust.\\ The ambiguity of the nature of the object at
the SN position in the pre-explosion and late-time images means that,
although SN~1999ev may have had an identified progenitor of some kind,
the previously derived progenitor properties are unreliable; even in
the interpretation that the source is a cluster, the reliance on
three-colour photometry leads to degeneracies in the reddening-age
solutions that prohibit a precise initial mass estimate for the
progenitor.

\subsection{SN~2003gd}
\label{sec:ana:03gd}
\begin{figure*}
\includegraphics[width=17.5cm]{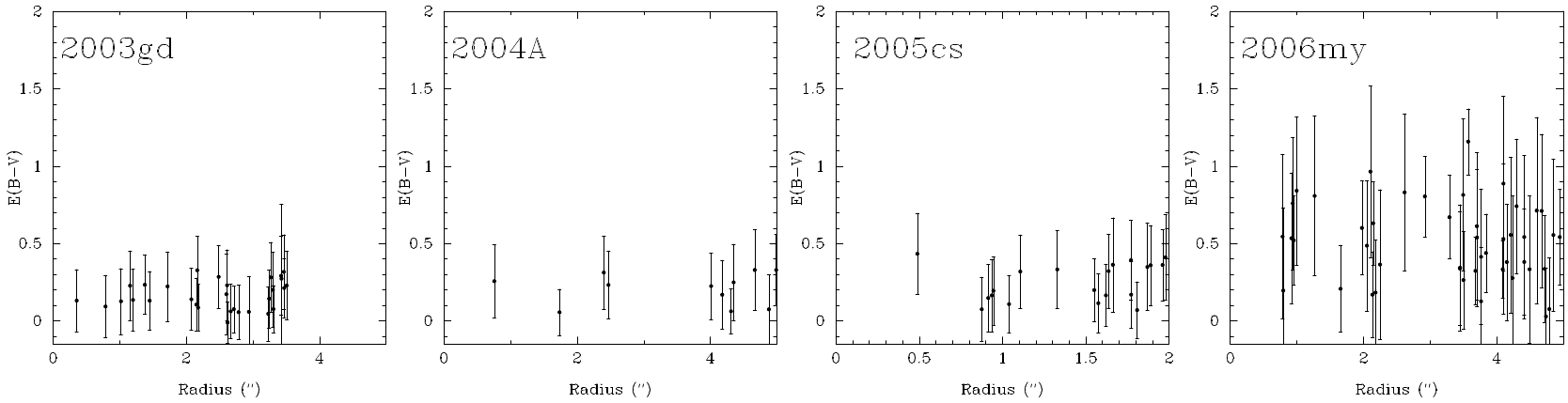}
\caption{Reddenings for stars surrounding the sites of SNe 2003gd, 2004A, 2005cs and 2006my (as a function of distance from the SN positions).}
\label{fig:ana:red}
\end{figure*}

The photometry of 30 stars within $4\arcsec$ ($\sim 200\mathrm{pc}$)
of the position of SN~2003gd was used to derive an average reddening
towards the SN site of $E(B-V)=0.14\pm0.04$ (see Figure
\ref{fig:ana:red}).  The amount of reddening is consistent with the
reddening previously estimated from three colour photometry of the
surrounding stars using early post-explosion $ACS HRC$ images
\citep{smartt03gd} and from the colour evolution of the SN itself
\citep{2005MNRAS.359..906H}.

\begin{figure*}
\includegraphics[width=5cm, angle = 270]{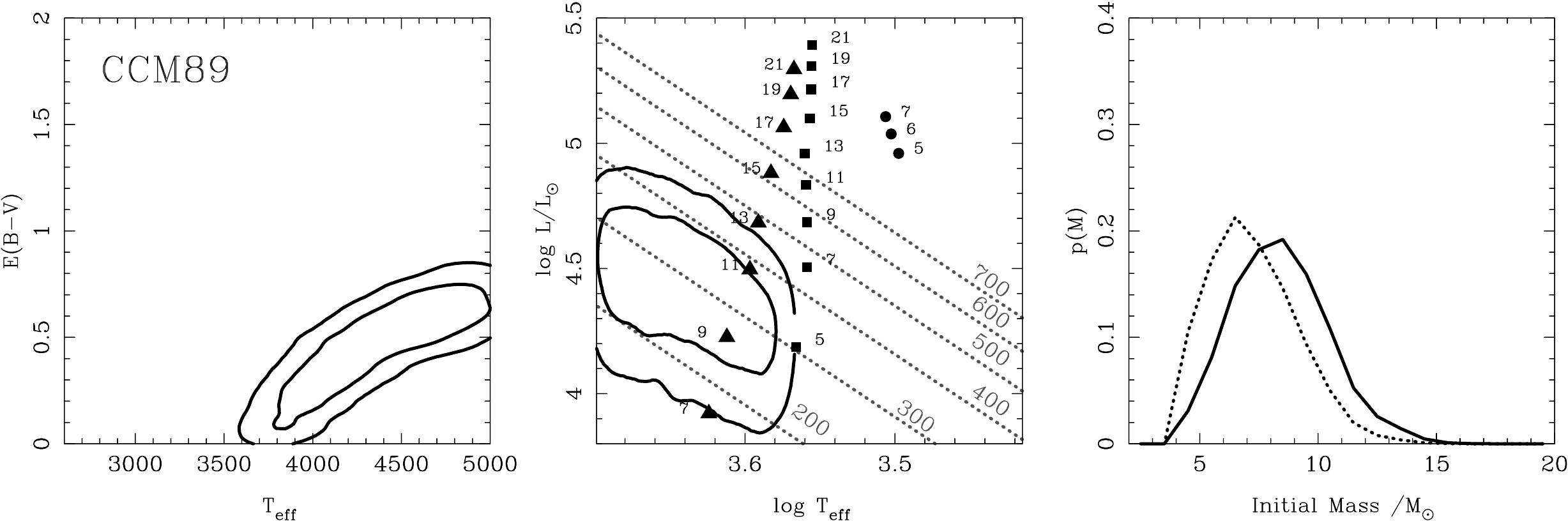}\\
\includegraphics[width=5cm, angle = 270]{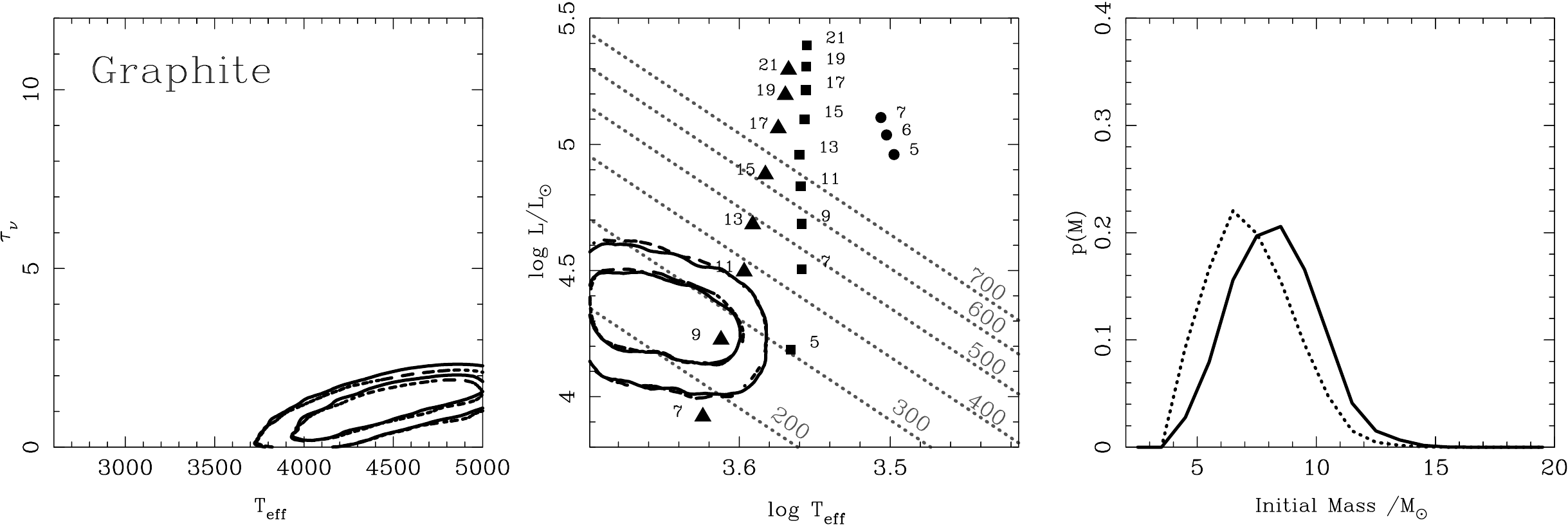}\\
\includegraphics[width=5cm, angle = 270]{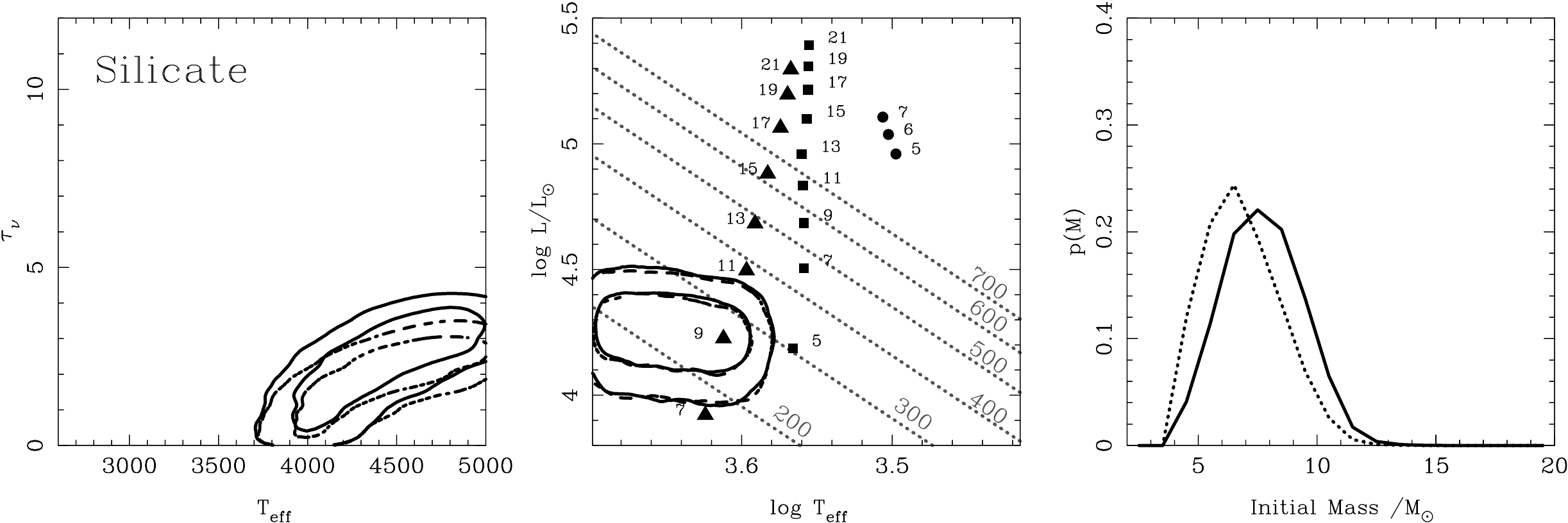}\\
\caption{The parameters of the progenitor of SN~2003gd for different
  reddening components.  In each panel the contours contain 68\% and
  95\% of the total probability.  {\it Top Row),Left)} The temperature
  and reddening of the progenitor assuming Galactic-like dust; {\it
    Centre)} the progenitor's location on the HR diagram (also shown
  are the locations of stellar evolution models of given initial mass
  for the end of core He burning ($\blacktriangle$), the onset of Ne
  burning ($\blacksquare$) and the endpoints for those models that
  undergo second dredge-up ($\bullet$).  Dotted grey lines indicate
  lines of constant progenitor radius.; and {\it Right)} the inferred
  initial mass probability density function with no weighting (solid
  line) and weighting according to the initial mass function (dotted
  line).  {\it Middle Row)} The same as the top row, but for a
  \citet{ccm89} reddening component of $E(B-V)=0.14\pm0.04$, derived
  from the surrounding stars and a unconstrained reddening component
  for graphite dust around the progenitor (solid contours are for
  $R_{out}/R_{in}=2$ and dotted contours are for $R_{out}/R_{in}=10$).
  The mass probability density function are as for the top row, but
  only shown for the $R_{out}/R_{in}=2$ solution.  {\it Bottom Row)}
  The same as the middle row, but for silicate dust. }
\label{fig:ana:03gdhrd}
\end{figure*}

Given the detection of the progenitor in pre-explosion observations in
two filters, we consider the roles of three different types of
reddening: 1) unconstrained reddening, with an interstellar reddening
law; 2) an interstellar reddening component consistent with the observed
reddening to the surrounding stars and an unconstrained degree of
reddening arising from Graphite dust around the progenitor; and 3) the
same as 2, except with Silicate dust.  The corresponding regions of
the parameter space and the Hertzsprung-Russell (HR) diagram, allowed
by the pre-explosion observations in conjunction with half-solar
metallicity MARCS SEDs, are presented on Fig. \ref{fig:ana:03gdhrd}.
Given the observed colour of the progenitor, regardless of the amount
of reddening, we find $T_{eff} > 3500K$ and the radius of the
progenitor constrained to be $200 <R < 400R_{\odot}$.  The flatter
nature of the graphite and silicate reddening laws, compared to the
\citeauthor{ccm89} reddening law, leads to tighter luminosity
constraints and a weaker dependence on the temperature/colour of the
progenitor.  The masses inferred for the progenitor are relatively
insensitive to the choice of reddening law, with $M_{init} \sim
7-8\pm2.0M_{\odot}$.

Given the reddening derived towards SN~2003gd, we find that the SED of
the object recovered at the SN position in the late-time images
($A^{\prime}$) is consistent with a black body with temperature
$T\sim9\,000\mathrm{K}$, which is similar to the expected temperature
for SNe at such late times \citep[see e.g][]{2012ApJ...744...26O}.

\subsection{SN~2004A}
\label{sec:ana:04A}
The three-colour photometry of 12  stars within 5.25$\arcsec$
($\approx 500\mathrm{pc}$; see Fig. \ref{fig:ana:red}) of the position
of SN2004A was used to derived a weighted average reddening towards the
SN site of $E(B-V)=0.16\pm0.06$.  This is larger than the estimate
made by \citet{2006MNRAS.369.1303H} of $E(B-V)=0.06\pm0.03$, derived
using three colour photometry from the post-explosion $ACS/WFC$
images.

\begin{figure*}
\includegraphics[width=5cm, angle = 270]{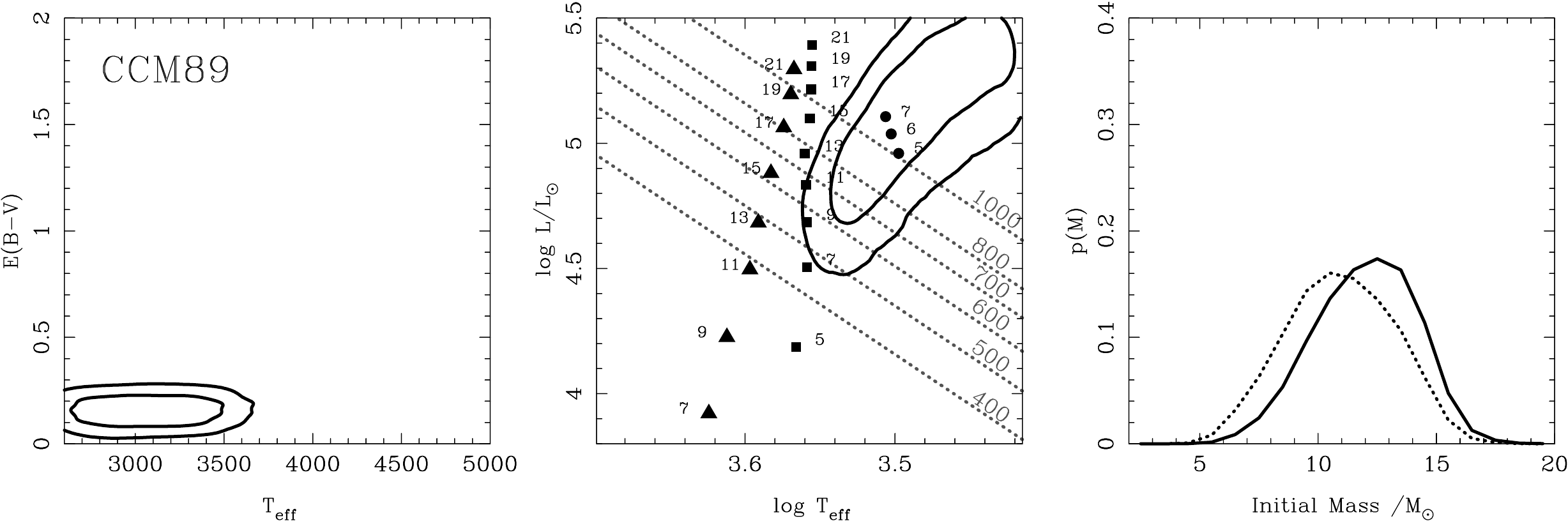}\\
\caption{The same as Fig. \ref{fig:ana:03gdhrd} but for pre-explosion
  observations of the progenitor of SN~2004A, assuming only a
  \citeauthor{ccm89} reddening law with $E(B-V)=0.06\pm0.03$ derived
  from surrounding stars.}
\label{fig:obs:04Ahrd}
\end{figure*}

Due to the single $F814W$ detection of the progenitor, and the upper
$F606W$ limit, an SED fit with $>1$ parameter is under-constrained by
the observations.  Using particular prior assumptions, such as the
reddening derived from surrounding stars, it is possible to probe the
allowed likely parameter space.  We conducted a two-parameter fit to
MARCS SEDs, with half-solar metallicity, for temperature and a single
reddening component (following a \citeauthor{ccm89} reddening law) with
 the reddening derived from surrounding stars as a prior.  The
resulting constraint in the temperature-reddening plane is shown on
Figure \ref{fig:obs:04Ahrd}.  The low reddening and the strict $F606W$ upper limit constrains the
temperature of the progenitor to be lower than $3700\mathrm{K}$;
although this is sensitive to the reddening prior, such that higher
reddenings may permit hotter progenitors.  The corresponding region on
the HR diagram is also shown on Figure \ref{fig:obs:04Ahrd}.  As the
reddening is effectively fixed, the slope of the contours reflects the
increasing bolometric correction for cooler stars.  If the reddening
inferred from the surrounding stars is, instead, a lower limit on the
reddening towards the progenitor (due to an additional component of
reddening due to circumsteller dust), then we can only derive a lower
limit on the luminosity of the progenitor.  The contours also include those stars in the initial mass range $5-7M_{\odot}$ that are expected, at this metallicity, to undergo second dredge up and become AGB stars.  
\subsection{SN~2005cs}
\label{sec:ana:05cs}
The photometry, from the pre-explosion images, of $20$ stars within
$2\arcsec$ ($\mathrm{\sim 120 pc}$) of the position of SN~2005cs were
used to determine a weighted average reddening towards the SN of
$E(B-V)=0.22\pm0.05$ (see Fig. \ref{fig:ana:red}).

\begin{figure*}
\includegraphics[width=5cm, angle = 270]{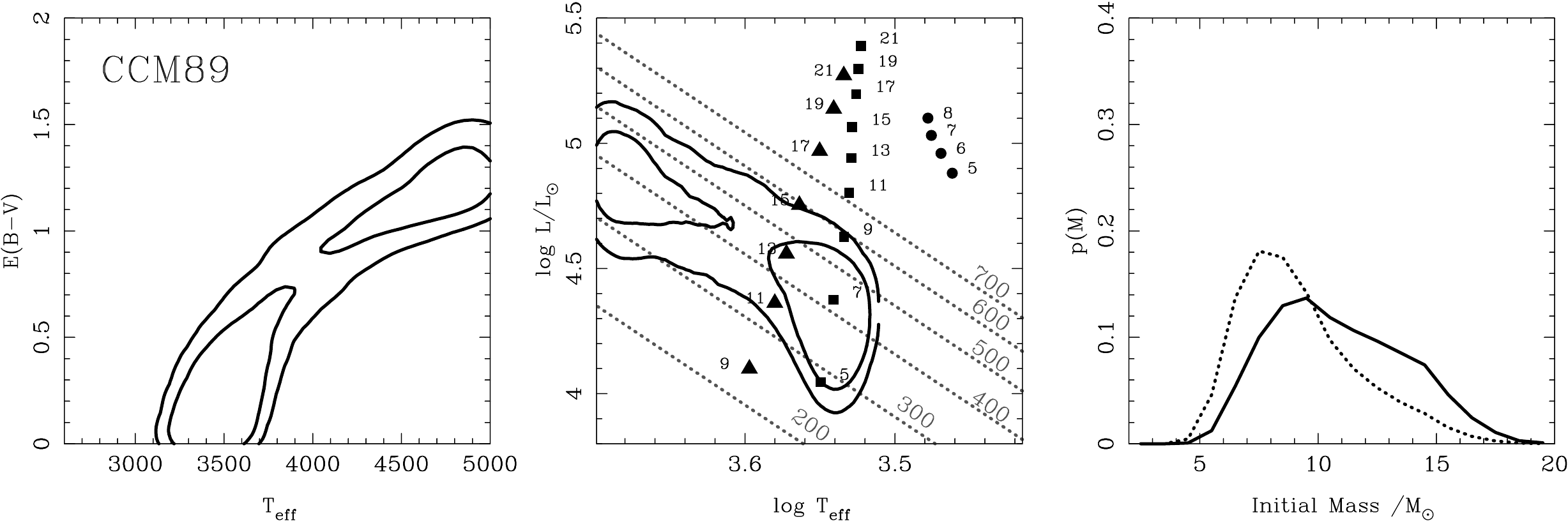}\\
\includegraphics[width=5cm, angle = 270]{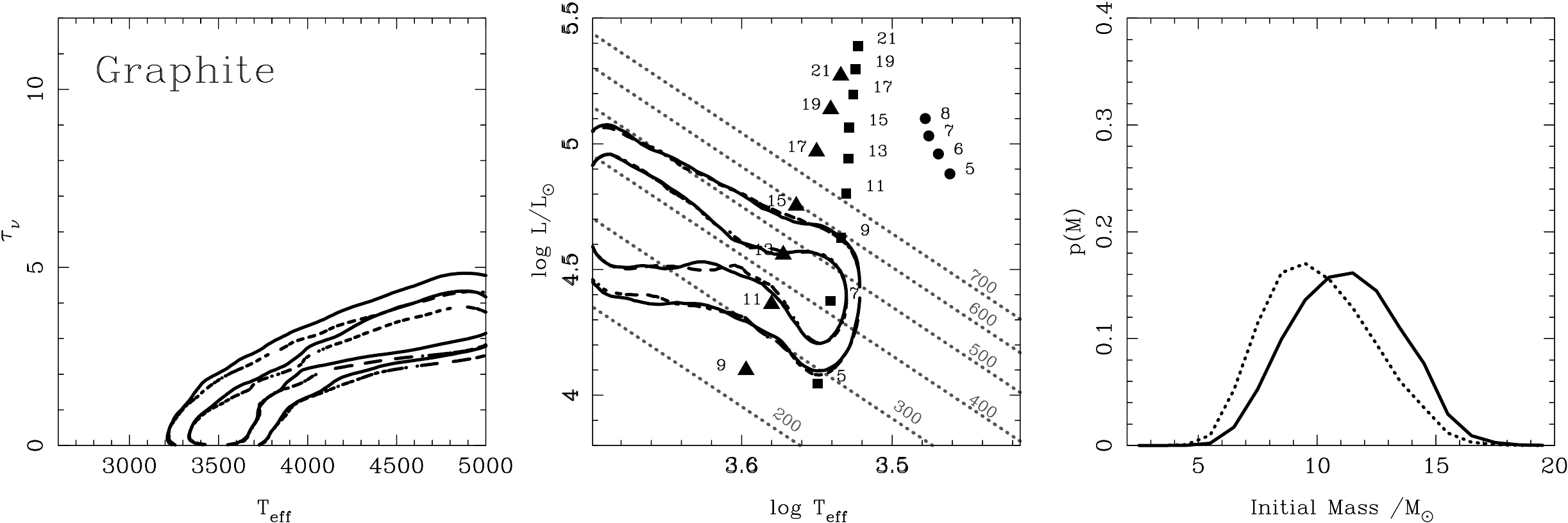}\\
\includegraphics[width=5cm, angle = 270]{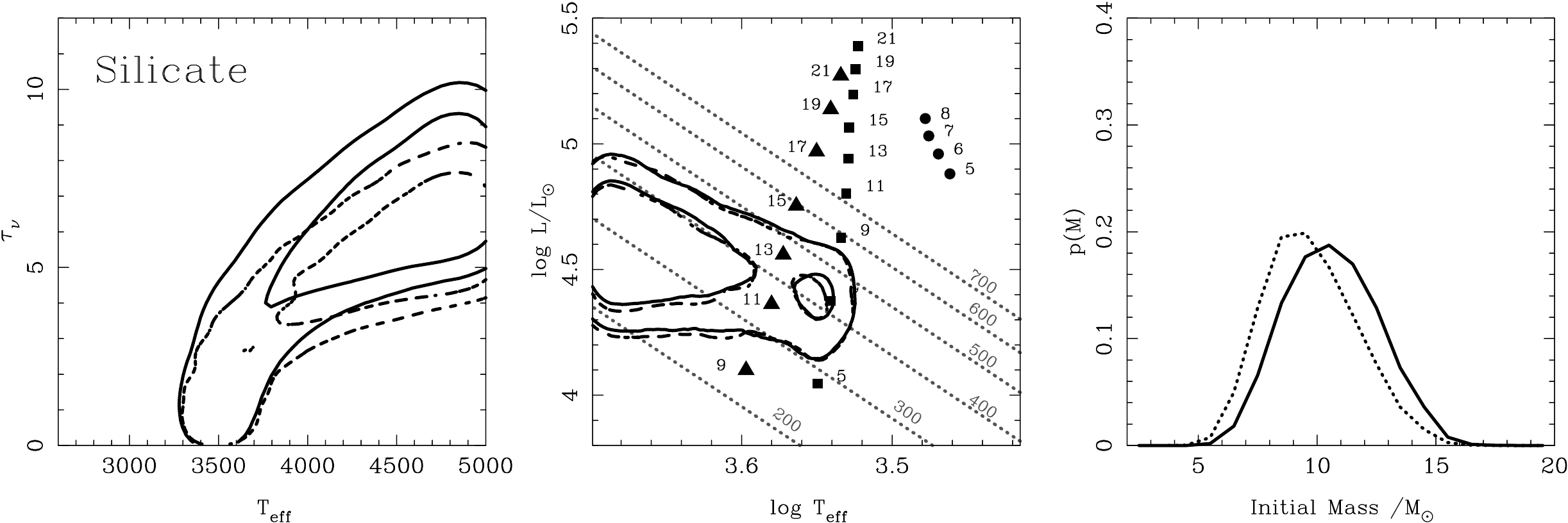}\\
\caption{The same as Fig. \ref{fig:ana:03gdhrd}, but for the pre-explosion observations of the progenitor of SN~2005cs.}
\label{fig:ana:05cshrd}
\end{figure*}

In order to constrain the properties of the progenitors we utilised
the $F814W$ magnitude presented here, measured using image
subtraction, and the revised detection thresholds for the
pre-explosion $ACS$ $F435W$, $F555W$ and $F658N$ images.  Furthermore,
we adopted the infrared detection limits for the progenitor reported
for pre-explosion Gemini $NIRI$ $JHK$ images
\citep{2005MNRAS.364L..33M} and $NICMOS$ $F110W$, $F160W$ $F222M$
images \citep{2005astro.ph..7394L}.  As the $NICMOS$ images are
deeper, they form the principal constraint on the SED of the
progenitor in the IR, however we include the Gemini $NIRI$ limits in
our calculation for completeness.  We explored the same parameter
space, for the same combinations of reddening components, as for the
progenitor of SN~2003gd (see Section \ref{sec:ana:03gd}) and the
results are presented on Fig. \ref{fig:ana:05cshrd}.

Similarly to the progenitor of SN~2003gd, the graphite and silicate
reddening laws yield flatter contours of the HR diagram, making the
dependence of the luminosity on the temperature less extreme.  For
each reddening type, there are two islands of preferred solutions: a
cool, low reddened solution and a hotter, reddened solution;
reflecting the severe constraints from non-detections in the infrared
and in blue, respectively.  The bimodal probability distribution in
$T_{eff}$ and $E(B-V)$ leads to a skewed mass probability density
function for a \citeauthor{ccm89} reddening law extending to higher
masses, although the peak of the distribution occurs at $\sim
10M_{\odot}$.  For the graphite and silicate reddening laws, the
unweighted mass probability density function is more symmetric and
peaks around $\sim 11M_{\odot}$.  The strict infrared limits exclude
the possibility of the progenitor being a massive AGB stars for all
the reddening types \citep{2007MNRAS.376L..52E}.

\subsection{SN~2006my}
\label{sec:ana:06my}
\begin{figure*}
\includegraphics[width=5cm, angle=270]{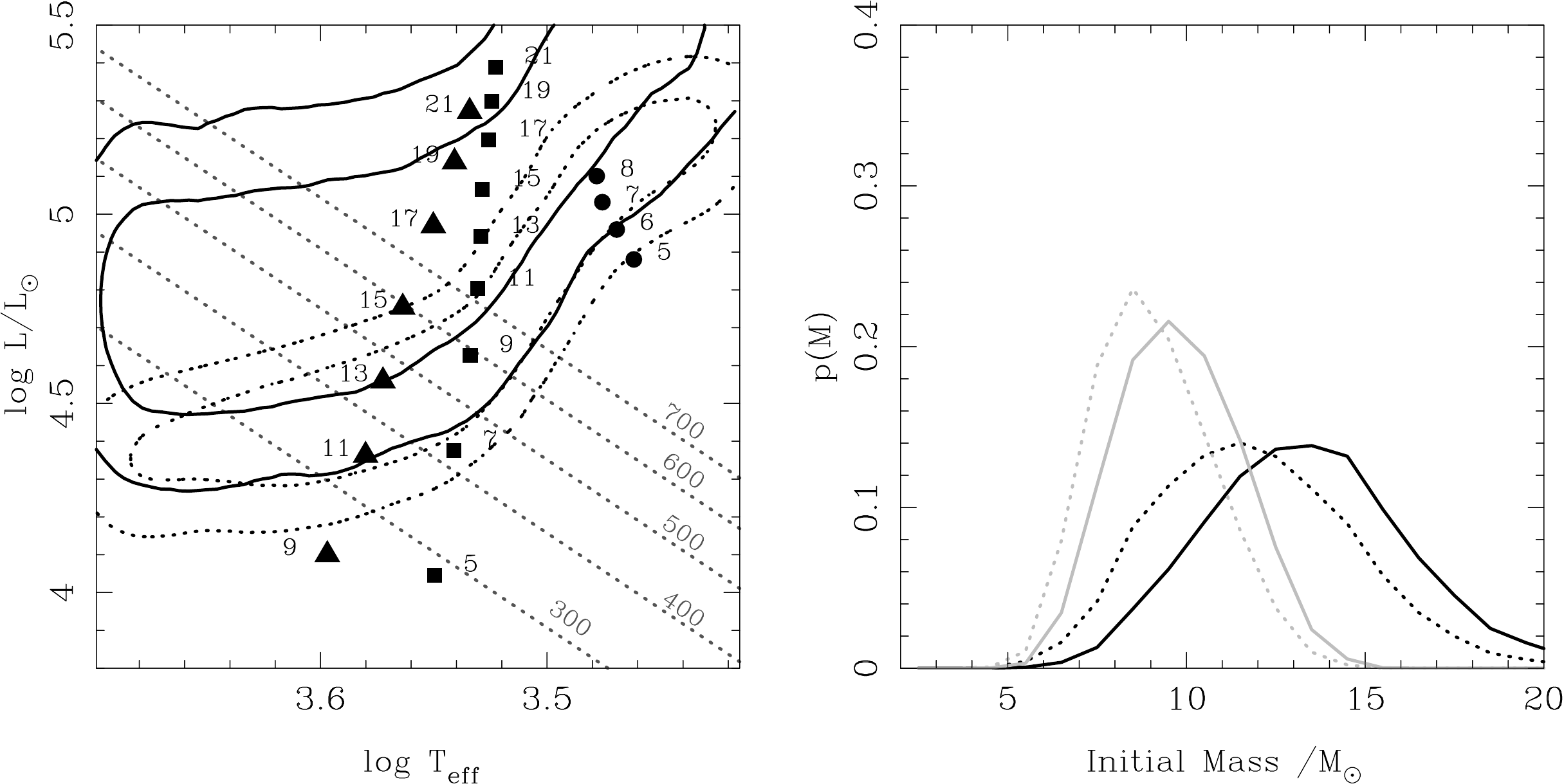}
\label{fig:ana:06myhrd}
\caption{The parameters of the progenitor of SN~2006my.  {\it Left)} The progenitor of SN~2006my on the HR diagram, assuming $E(B-V)=0.49\pm0.25$ (solid contours) and $E(B-V)=0.027$ (dotted contours). {\it Right)}  The initial mass probability density functions for the progenitor of SN~2006my, given $E(B-V)=0.49\pm0.25$ (heavy curves) and $E(B-V)=0.027$ (grey curves).}
\end{figure*}

Previously \citet{2007ApJ...661.1013L} suggested that, apart from a
Galactic reddening component, there was no evidence for a host
reddening component in spectra of SN~2006my; this value was similarly
used by progenitor studies conducted by
\citet{2011MNRAS.410.2767C} and \citet{2008PASP..120.1259L}.  We
utilised the HSTphot photometry of the post-explosion WFPC2
images of SN~2006my to study the reddening associated with the
surrounding stellar population.  We selected 46 good stars
($\chi^{2}<1.5$, $\mathrm{|sharp|} < 0.3$) with complete three-colour
photometry within $5\arcsec$ ($\mathrm{\sim 540 pc}$) of the SN.  This
photometry was compared with solar metallicity supergiant SEDs, to
derive an average reddening towards SN~2006my of $E(B-V)=0.49\pm0.25$
(see Fig. \ref{fig:ana:red}).  The large scatter reflects the poor
quality of the {\it WFPC2} photometry, as well as likely differences
in the reddening between the individual stars.  The relative colour
constraint provided by the pre-explosion $F555W$ limit and the $F814W$
detection of the progenitor is insufficient to limit the temperature
of the progenitor for temperatures below $< {4500K}$.  The
allowed regions of the HR diagram, for the two reddening estimates, is
shown on Figure \ref{fig:ana:06myhrd}.  As for SN~2004A (see Section
\ref{sec:ana:04A}), the shape of the contours is dictated by the larger
bolometric correction at cooler temperatures.  The lack of an infrared
detection or detection limit for the progenitor means that both the
RSG and AGB solutions are allowed.  The higher reddening inferred from
the photometry of the surrounding stars implies higher luminosities
for the progenitor, than for just Galactic foreground reddening, and leads to
a higher initial mass ($13.4\pm2.8$ vs. $9.8\pm1.7M_{\odot}$).

\begin{figure*}
\includegraphics[width=18cm]{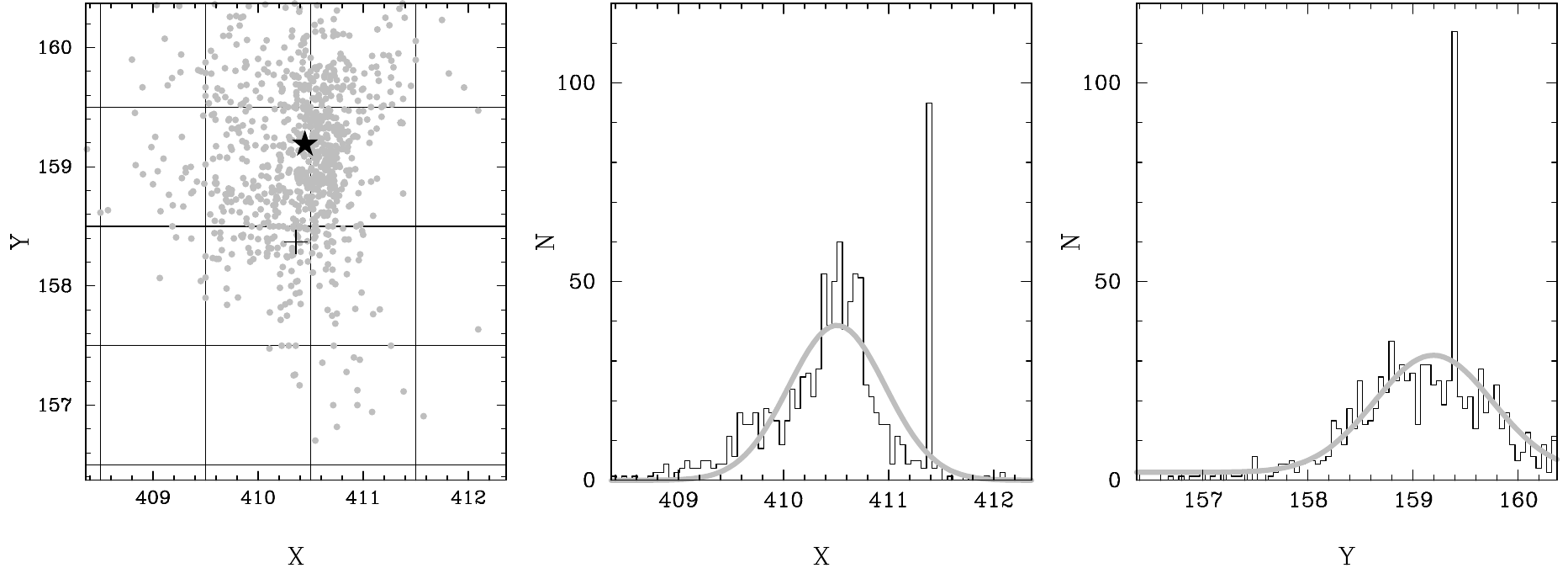}
\caption{The results of position determinations for the pre-explosion source, using  DAOphot and the ofilter centring algorithm, for 1000 Monte Carlo simulations of the pre-explosion $F814W$ image of the site of SN~2006my.  {\it Left)} The positions of the recovered sources in the coordinate frame of the original $WFPC2$ image (in the DAOphot coordinate system).  The position of the source recovered by HSTphot is indicated by the $+$, while the mean position derived form the Monte Carlo simulations is indicated by the $\star$; {\it Centre and Right)} The position distributions in $x$ and $y$ coordinates for the recovered sources; the grey line indicates a Gaussian fit to the distributions.} 
\label{fig:ana:06mycent}
\end{figure*}

Given the apparent offset between the transformed SN position and the
source recovered in the pre-explosion $F814W$ image, we explored the
possible causes for the apparent discrepancy.  We note that all three
previous studies, and our own results, agree that there is a
significant offset between the transformed SN position and the
pre-explosion $F814W$ source.  As noted by
\citet{2008PASP..120.1259L}, in conjunction with \citet{dolphhstphot},
the positional uncertainty for an isolated source with the brightness
of the pre-explosion $F814W$ source is $\sim 0.4$ pixels.  We
conducted Monte Carlo simulations, where the pre-explosion image was
``resampled'' under the assumption that the observed counts could be
modelled as source and background flux (following a Poisson
distribution) with a Gaussian readnoise contribution.  Aperture photometry of
these simulated images was conducted using DAOphot, with the
ofilter centring algorithm, to recover the position of the source
(within a 5 pixel centring box).  The result of these simulations is
shown on Fig. \ref{fig:ana:06mycent} and the average position is
$x=410.5\pm0.5$, $y=159.2\pm0.6$; offset from the SN position by 0.4
pixels.  From the outcome of the Monte Carlo simulations we note three
effects:
\begin{enumerate}
\item{The positional uncertainty is larger for faint sources and will be dependent on pixel noise statistics in the main pixel containing source flux and the surrounding pixels.}
\item{Given the subsampled nature of the PSF in $WFPC2$ images, the uncertainty on the position of a source will also be dependent on the brightness of the immediately surrounding pixels; and, as evident from Fig. \ref{fig:ana:06mycent}, the measured position may be skewed by the proximity of the true source position to a pixel edge.}
\item{The position derived is sensitive to the choice of centring algorithm utilised.  Trials using the ``centroid'' centring algorithm in DAOphot showed that positions were preferentially located at integer and half-integer pixel coordinates.}
\end{enumerate}
We conclude that, in addition to uncertainties associated with determining positions on {\it WFPC2} images for isolated sources, it is also important to consider the effects of nearby sources (within a few pixels) that may skew/bias the centring algorithm away from the true source position.  From the Monte Carlo simulations, given the environment and pixel noise at the SN position, we estimate the uncertainty on the position of the pre-explosion $F814W$ source may be $\sim0.078\arcsec$.  With such large uncertainties, the apparent offset between the SN and the pre-explosion source would be only $\sim 1 \sigma$ given the offsets calculated by \citet{2011MNRAS.410.2767C} and \citet{2008PASP..120.1259L}; however, we caution that our lower quality geometric transformation results in a $1.2\sigma$ offset.

\section{Discussion}
\label{sec:disc}
A summary of the masses derived for the progenitors considered here is
presented on Table \ref{tab:disc:mass}.  The biggest differences
between our findings and those of \citet{2008arXiv0809.0403S} are for
the progenitors of SNe~1999ev and 2006my.  Our late-imaging shows that
the nature of the source found at the SN position in pre-explosion
observations of the site of SN~1999ev is, at best, uncertain; whilst,
at worst, a misidentification of a host cluster or unrelated coincident
star.  As the nature of this source is not clarified in the late-time
imaging, the progenitor of SN~1999ev should no longer be considered in
progenitor population statistics.  In the analysis presented by
\citet{2008arXiv0809.0403S}, the progenitor of SN~1999ev had the
distinction of having the highest mass inferred for a detected Type IIP SN
progenitor, and so set the maximum mass limit for stars
to explode as RSGs and produce Type IIP SNe.  With the
removal of the SN 1999ev progenitor from the population statistics,
the maximum mass limit will actually drop and make the ``Red
Supergiant Problem'' even more severe.  The late-time imaging of the
site of SN~2006my has shown that the progenitor was detected in the
pre-explosion observations, and the upper mass limit quoted by
\citeauthor{2008arXiv0809.0403S} should now be quoted as a detected
progenitor with a corresponding mass estimate.

The mass estimates derived for the confirmed progenitors are generally
higher (by $\sim 1M_{\odot}$) than those presented by
\citeauthor{2008arXiv0809.0403S}; in part, due to the slightly larger
foreground and host reddenings we inferred towards the progenitors
from the colours of the surrounding stars.  In addition, for SN~2003gd
and 2005cs we considered additional reddening components to the
reddening from interstellar dust, in keeping with the expectation that
there is dust local to progenitor that is not probed by the
surrounding stars or the observations of the SNe.  We also note that
our uncertainties are smaller than those quoted by
\citeauthor{2008arXiv0809.0403S}: we considered the effects of
uncertainties of the luminosity convolved with the flat probability
distribution of the star having a mass in the range bounded by the
maximum mass star to end core He-burning at that luminosity and the
minimum mass star to begin Ne burning at that luminosity.  The initial
mass probability density functions are, apart from SN~2005cs,
approximately symmetric and almost follow a normal distribution.  We
believe this is a fairer presentation of the initial masses for the
progenitors and their uncertainties.  The application of a weighting
to the initial mass pdf, according to the IMF, has a small effect in
shifting the pdf to slightly lower masses.  Unlike
\citet{2008arXiv0809.0403S}, who used this weighting scheme to
``truncate'' their large uncertainties, we find that the effect of
weighting on the relative width of the initial mass pdfs is minor.

The slight increase in the inferred progenitor mass does not help
rectify the apparent discrepancy between these ``evolutionary'' masses
and the progenitor masses derived from hydrodynamical models.
\citet{2008A&A...491..507U} found an initial mass for the progenitor
of SN~2005cs of $18.2\pm1M_{\odot}$, which is at odds with the new
masses derived here, regardless of the choice of reddening.  As
\citet{2008A&A...491..507U} suggest, additional dust in a
circumstellar shell could lower the apparent luminosity of the
progenitor and decrease the mass inferred from pre-explosion
observations.  The nature of the pre-explosion observations of
SN~2005cs, in particular the strict near-infrared upper limits,
severely limits the amount of reddening that the progenitor might
undergo.  The constraints on the radius for the progenitor of
SN~2005cs are also below the radius inferred by
\citet{2008A&A...491..507U} of $600\pm140R_{\odot}$, but not
significantly discrepant.

It is clear from the analysis presented in Section \ref{sec:ana}, that
the available pre-explosion observations of a given SN progenitor
dictates the degree of analysis that may be conducted.  Given the
presence of constraining optical and infrared upper limits, the
possible effect of circumstellar reddening on the progenitor of
SN~2005cs was evaluated despite having a detection in only one band.
This demonstrates the importance for having good pre-explosion
observations in the near-infrared for studying the cool progenitors of
Type IIP SNe, even if the progenitor is not detected at those
wavelengths.  The two detections of the progenitor of SN~2003gd, at
different wavelengths, enabled similar constraints for reddening due
to circumstellar dust.  Conversely, the analyses of SN~2004A and
2006my were limited by them having only a single detection and only
loose constraint on the progenitor colour at a bluer wavelength.  

Due to the lack of constraints on reddening due to circumstellar dust,
the final derived masses for the progenitors of SNe 2004A and 2006my
may represent, in actuality, lower mass limits.  For this study, we
have assumed the circumstellar dust follows the reddenings laws
proposed by \citet{2012ApJ...759...20K}, for the case of the
progenitor of SN~2012aw.  Conversely, \citet{2012ApJ...756..131V}
suggested a steeper reddening law, still following \citeauthor{ccm89},
but with $R_{V}\approx 4.35$; leading to an overall higher extinction.
The \citet{2012ApJ...759...20K} reddening law was determined using
models of specific dust compositions, of either graphite or silicate
dust, expected to be found around RSGs, whereas
\citet{2012ApJ...756..131V} estimated the change in reddening law
based on observations of Galactic RSGs.  For SN~2012aw
\citep{2012arXiv1204.1523F,2012ApJ...756..131V}, the progenitor was
detected in four bands, and the degeneracies between reddening,
reddening law and temperature could not be broken; suggesting the
full determination of the parameters, independent of the
assumptions of reddening laws, requires detections at $>4$
wavelengths, such as for the progenitor of SN~2008bk
\citep{2008ApJ...688L..91M,2012AJ....143...19V}.

Both \citet{2012arXiv1204.1523F} and \citet{2012ApJ...756..131V} noted
the large decrease in reddening determined for the pre-explosion
source and for the subsequent SN; suggesting such large reddenings may
affect all progenitors but not be apparent post-explosion.  For their
sample of RSGs, \citet{2013arXiv1302.2674D} measure extinctions
arising from circumstellar dust in the range $A_{V}=0.0-1.0$.  A
further issue, that we have not explored, was suggested by
\citet[][and references therein]{2012MNRAS.419.2054W} that the amount
of dust in the circumstellar medium is related to the mass loss rate
of the RSG and, ultimately, its bolometric luminosity.

A corresponding issue to the reddening problem is the temperature.  We
note that, for the progenitors with constraining pre-explosion
observations, we find the allowed temperature range to be generally
hotter than the predicted endpoints for the stellar evolution models,
but are consistent with the recent reappraisal of RSG temperatures by
\citet{2013arXiv1302.2674D}.  The lower limit of the temperature
scales for the progenitors of SNe~2003gd and 2005cs suggests that they
have a spectral type no later than M0, which corresponds to the
predicted positions for stars that have just finished core He-burning;
the pre-explosion observations of SNe~2003gd and 2005cs suggest that
the progenitors were not massive AGB stars
\citep{2007MNRAS.376L..52E,2007A&A...476..893S}.  It is only in the
poorly constrained cases, for the progenitors of SNe~2004A and 2006my, that
we cannot exclude cooler temperatures that might be associated with
massive AGB stars.  With limited observations in the optical (in
particular the $B$ and $V$ bands), it is difficult to place limits on
the maximum temperature of the progenitor, as hotter temperatures can
always be accommodated with additional reddening.  In the case of the
progenitor of SN~2005cs, the requirement that the progenitor ended its
life as an RSG has serious implications for the interpretation of the
subsequent SN as a low luminosity ``Electron-Capture'' SN
\citep{2012ARNPS..62..407J}.

Our late-time imaging campaign has shown that, for the case of
SN~2003gd, the possibilities of recovering precise photometry of the
progenitor through template subtraction may be undermined by
rebrightening of the SN at late-times.  In the case of SN~2003gd, the
previous analysis of \citet{2009Sci...324..486M} was fortuitous in
that it managed to observe the SN before it rebrightened with a very
strict brightness limit of the SN in the late-time $i^{\prime}$ image.
Such rebrightening is not without precedent;
\citet{2009ApJ...704..306K} observed the optical lightcurve of
SN~2004et to rebrighten (by $\sim 1$ mag in $V$) at optical and
infrared wavelengths.

As noted in Section \ref{sec:ana:06my}, the apparent discrepancy
between the position of the progenitor of SN~2006my on the
pre-explosion images and the transformed SN position does raise
questions about how positional uncertainties are handled.
\citet{2005astro.ph..1323M} determined the positional precision using
the standard deviation of the four different centring methods
available to DAOphot (centroid, ofilter, gauss and psf).  It must be
noted, however, that these centring techniques are not providing {\it
  independent} estimates of an object's location.  In the future, it
may be preferable to choose a single centering technique, and consider
the role of Poisson noise (both object and background) and read out
noise in each pixel on the determination of an object's position.
Furthermore, tests using the centroiding routines in both the DAOphot
and SAO image DS9 packages has shown that, in the case of isolated
objects in subsampled images, centroiding can be a relatively blunt
tool (providing default positions located in the centre of pixels).
In considering flux deficits using image subtraction techniques,
rather than appealing to astrometric coincidence, we have shown that
confirmation of a star as being the actual progenitor requires
observing it to have disappeared.
 
\begin{table*}
\caption{\label{tab:disc:mass}  Final results for the progenitors of SNe~1999ev, 2003gd, 2004A, 2005cs and 2006my, for different reddenings due to instellar dust (CCM99) and circumstellar dust reddening laws (CSM Graphite and CSM Silicate).}
\begin{tabular}{lcccccccc}
\hline\hline
SN            &   \multicolumn{2}{c}{CCM89}  & & \multicolumn{2}{c}{CSM Graphite}  & & \multicolumn{2}{c}{CSM Silicate}  \\
\cline{2-3} \cline{5-6}\cline{8-9} \\
            &   w/ IMF      & w/o IMF     & &    w/IMF        &  w/o IMF  &   &    w/  IMF      & w/o IMF              \\  
\hline
1999ev      &  \multicolumn{2}{c}{Likely cluster} & & $\cdots$ & $\cdots$    & &  $\cdots$       & $\cdots$             \\
2003gd      &  $7.3\pm1.9$    &   $  8.4\pm 2.0$ &   & $7.3\pm 1.8$   & $ 8.2\pm 1.8$ & &  $6.9\pm1.6$   & $7.7\pm1.6$       \\  
2004A       &  $10.9\pm2.3$   &   $ 12.0\pm2.1$ & & $\cdots$ & $\cdots$ & & $\cdots$ & $\cdots$   \\   
2005cs      &  $7.9^{+2.6}_{-1.6}$$^{a}$  &$9.5^{+3.4}_{-2.2}$$^{a}$ & &  $10.1\pm2.2$ & $11.2\pm 2.2$ & &$9.7\pm1.9$ & $10.6\pm 1.9$ \\
2006my$^{b}$ & $9.1\pm1.7$    & $9.8\pm1.7$ &  &$\cdots$ & $\cdots$     &  &$\cdots$       & $\cdots$             \\
2006my$^{c}$ & $11.9\pm2.7$  & $13.4\pm2.8$ &  &$\cdots$ & $\cdots$     &  &$\cdots$       & $\cdots$             \\
\hline\hline    
\\
\multicolumn{7}{l}{$^{a}\,$ The values corresponding the mode and $68\%$ probability intervals.}\\
\multicolumn{7}{l}{$^{b}\,$ Assuming foreground reddening $E(B-V)=0.027$.}\\
\multicolumn{7}{l}{$^{c}\,$ Assuming $E(B-V)=0.49\pm0.25$.}\\
\end{tabular}
\end{table*}

\section{Conclusions}
We have presented late-time imaging of the sites of five Type IIP SNe
with pre-explosion HST images, in which progenitor candidates were
detected.  In three of the cases (2003gd, 2004A and 2005cs), our
previous identifications have been confirmed and we find initial masses for these
stars in the range $6-14M_{\odot}$.  The pre-explosion observations of
SNe 2003gd and 2005cs are sufficient to place constraints on the
progenitor mass that are relatively insensitive to the amount and type
of dust around these progenitors.  Given the similarities in
brightness between the pre-explosion and late-time sources detected at
the position of SN~1999ev, we conclude the progenitor identification
for this SN is unsafe and suggest the pre-explosion source may be a
reddened host cluster; although the three-colour late-time imaging is
insufficient to place a tight constraint on the age or reddening of a such
cluster.  The analysis of the pre-explosion and late-time observations
of the site of SN~2006my have revealed that the source previously
thought to be significantly offset from the SN position has
disappeared.  The astrometric coincidence of the residual in the
difference image with the transformed SN position suggests it was the
progenitor object.

Far from providing just simple confirmation of a progenitor's identity
(through its disappearance), our analysis shows late-time imaging is crucial for
conducting a deeper and more precise analysis of the properties of a
progenitor than is afforded by fortuitous pre-explosion observations
alone.  The power of the application of late-time imaging, for
studying progenitors, is demonstrated by the significantly deeper
detection limits that may be achieved by using artificial star tests
in conjunction with image subtraction techniques.

\section*{Acknowledgements}
Based on observations made with the NASA/ESA Hubble Space Telescope,
which is operated by the Association of Universities for Research in
Astronomy, Inc., under NASA contract NAS 5-26555. These observations
are associated with program GO-11675.  The research of JRM is funded
through a Royal Society University Research Fellowship.  We thank Stephen Smartt and Steen Hansen for their useful comments.

\bibliographystyle{mn2e}

\end{document}